\documentclass[10pt,journal,compsoc]{IEEEtran}
\ifCLASSOPTIONcompsoc
  \usepackage[nocompress]{cite}
\else
  \usepackage{cite}
\fi
\ifCLASSINFOpdf
\else
\fi
\usepackage{amsfonts}
\usepackage{graphicx}

\usepackage{algorithm}

\usepackage{algorithmic}

\usepackage{amssymb}
\usepackage{mathrsfs}
\usepackage{mathtools}
\usepackage{tikz-cd}
\usepackage{amsmath}
\usepackage{enumerate}
\usepackage{comment}
\usepackage{enumitem}
 \let\oldcdot\cdot
 \usepackage{breqn}
 \let\cdot\oldcdot

\usepackage{amsthm}
\usepackage{caption}
\usepackage{multirow}
\usepackage{lineno}
\DeclareMathAlphabet{\mathcal}{OMS}{cmsy}{m}{n}

\newtheorem{theorem}{Theorem}[section]
\newtheorem{lemma}{Lemma}[section]

\renewcommand{\tilde}{\widetilde}

\newcommand{\figref}[1]{Figure \ref{#1}}

\newcommand{\secref}[1]{Section~\ref{#1}}

\newcommand{\eqnref}[1]{equation (\ref{#1})}

\newcommand{\thmref}[1]{Theorem \ref{#1}}
\newcommand{\algoref}[1]{Algorithm \ref{#1}}
\newcommand{\etal}{et~al.}

\newcommand{\RG}{\mathcal{RG}}
\newcommand{\RS}{\mathcal{RS}}
\newcommand{\JCN}{\mathrm{JCN}}
\newcommand{\MDRG}{\mathbb{MR}}

\newcommand{\x}{\mathbf{x}}

\newcommand{\y}{\mathbf{y}}

\newcommand{\bv}{\mathbf{v}}

\newcommand{\f}{\mathbf{f}}

\newcommand{\g}{\mathbf{g}}
\newcommand{\h}{\mathbf{h}}

\newcommand{\ux}{\underline{x}}

\newcommand{\M}{\mathbb{M}}
\newcommand{\cM}{\mathcal{M}}
\newcommand{\R}{\mathbb{R}}

\newcommand{\Z}{\mathbb{Z}}

\newcommand{\PD}{\mathcal{PD}}
\newcommand{\Dg}{\mathit{Dg}}
\newcommand{\ExDg}{\mathit{ExDg}}

\newcommand{\cS}{\mathbb{S}}
\newcommand{\pI}{\mathit{pI}}

\newcommand{\pointa}{\mathbf{p_1}}
\newcommand{\pointb}{\mathbf{p_2}}
\newcommand{\cO}{\mathcal{O}}
\newcommand{\abs}[1]{\lvert{#1}\rvert}

\usepackage{array}

\begin{document}
%

\title{A Topological Distance between Multi-fields based on Multi-Dimensional Persistence Diagrams}
%
%
%
%

\author{}

\author{Yashwanth~Ramamurthi and~Amit~Chattopadhyay,~\IEEEmembership{Member,~IEEE}
\IEEEcompsocitemizethanks{\IEEEcompsocthanksitem Y. Ramamurthi and A. Chattopadhyay are in International Institute of Information Technology (IIIT), Bangalore.\protect\\
E-mail: \{yashwanth, a.chattopadhyay\}@iiitb.ac.in}
}
\IEEEtitleabstractindextext{%
\begin{abstract}
The problem of computing topological distance between two scalar fields based on Reeb graphs or contour trees has been studied and applied successfully to various problems in topological shape matching, data analysis, and visualization. However, generalizing such results for  computing   distance measures between two multi-fields based on their Reeb spaces is still in its infancy. Towards this, in the current paper we propose a technique to compute an effective distance measure between two multi-fields by computing a novel \emph{multi-dimensional persistence diagram} (MDPD) corresponding to each of the  (quantized) Reeb spaces. First, we construct a multi-dimensional Reeb graph (MDRG), which is a hierarchical decomposition of the Reeb space into a collection of Reeb graphs. The MDPD corresponding to each MDRG is then computed based on the persistence diagrams of the component Reeb graphs of the MDRG. Our distance measure extends the Wasserstein distance between two persistence diagrams of Reeb graphs to MDPDs of MDRGs. We prove that the proposed measure is a pseudo-metric and satisfies a stability property. Effectiveness of the proposed distance measure has been demonstrated in (i) shape retrieval contest data - SHREC $2010$ and (ii) Pt-CO bond detection data from computational chemistry.  Experimental results show that the proposed distance measure based on the Reeb spaces has more discriminating power in clustering the shapes and detecting the formation of a stable Pt-CO bond as compared to the similar measures between Reeb graphs.
\end{abstract}

}

\maketitle

\IEEEdisplaynontitleabstractindextext

%
\IEEEpeerreviewmaketitle

\IEEEraisesectionheading{\section{Introduction}\label{sec:intro}}

%
%
%
%

\IEEEPARstart{C}{omputing} distance or similarity between a pair of shapes or data is an important problem in topological data analysis. The problem of computing distance measures using scalar topology has been studied extensively and proven useful in shape or data clustering, symmetry detection, and feature extraction~\cite{hilaga2001topology, Zhang2004FastMO,2002-Edelsbrunner-Persistence,2023-Sridharamurthy-Comparitive-Analysis-of-Merge-Trees}. However, not all features in a data can be described using the scalar topology. Therefore, developing new tools for computing multivariate or multi-field (consisting of multiple scalar fields) topology is indispensable~\cite{2012-Duke-VisWeek,2016-MultiscaleMapper}. The Reeb space is one such topological structure that generalizes the Reeb graph, by generalizing the level set topology of a scalar field to capture the fiber topology of a multi-field \cite{2008-edels-reebspace}.  

Topological distances or similarities based on Reeb graphs, contour trees, and merge trees have been studied in the literature \cite{hilaga2001topology,Zhang2004FastMO,2022-Point-Wasserstein-Distance-Merge-Trees, 2022-Wetzels-Edit-Distance-Merge-Trees}. The persistence diagram is another important structure to capture the  homological changes in the data under a filtration ~\cite{2002-Edelsbrunner-Persistence}. Bottleneck distance and its generalization Wasserstein distance are two topological distances between persistence diagrams which have been applied in various problems of topological data analysis and machine learning~\cite{2014-Li-Persistence-Based-Structural-Recognition,2021-Rieck}.
Recently, a stable functional distortion metric  has been proposed between two Reeb graphs by Bauer \etal~\cite{2014-bauer-ReebGraph}. In the same paper, a bottleneck distance between two Reeb graphs has been introduced based on their persistence diagrams. Moreover, it is known that the bottleneck distance is insensitive to the details of bijections between the persistence diagrams, whereas the Wasserstein distance is more appropriate to handle that~\cite{book-herbert-computational-topology}.
However, extending such measures between two Reeb spaces is  challenging as we require a multifiltration to describe the  homological features of a Reeb space and no persistence diagram is known corresponding to a multifiltration. 
Towards this, our main contributions in the current paper are as follows: 
\begin{itemize}
    \item \textbf{A multi-dimensional persistence diagram representation corresponding to a quantized Reeb space:} In \secref{sec:persistent-mdrg}, we first decompose a quantized Reeb space or Joint Contour Net (JCN) into a multi-dimensional Reeb graph (MDRG). Then corresponding to each MDRG we define a \emph{multi-dimensional persistence diagram} (MDPD) to capture the persistent features of the component  Reeb graphs in the MDRG. 
    
    \item \textbf{A distance measure between two quantized Reeb spaces:} Next, using the MDPD representation, we define a distance measure between two quantized Reeb spaces. Our distance is an extension of the distance between persistence diagrams of two Reeb graphs \cite{2014-bauer-ReebGraph}. 
    
    \item \textbf{Properties of the distance measure:} We show that the proposed distance measure is a pseudo-metric and satisfies a stability property.

    \item \textbf{Complexity analysis:}
    We show the time complexity of computing the MDPD from a JCN and the complexity of computing the distance between two JCNs.
    
    \item \textbf{Applications in shape matching and data analysis:} In \secref{sec:experiments}, we show the effectiveness of the proposed distance measure in two different domains:
    (i) A shape contest data SHREC $2010$ - for clustering shapes and
    (ii) A Density Functional Theory (DFT) data from computational chemistry - for the detection of stable bond formation between Pt and CO molecules.
\end{itemize}

The paper is organized as follows. In \secref{sec:related-work}, we discuss the related works. \secref{sec:background} describes the background to  understand our proposed method. The main algorithm for computing the distance measure between two multi-fields, its properties, and the complexity analysis are discussed in \secref{sec:persistent-mdrg}.  
We show the  experimental results using the proposed distance measure in \secref{sec:experiments}, which is followed by a conclusion in \secref{sec:conclusions}.

\section{Related Work}
\label{sec:related-work}
Various similarity and distance measures between scalar fields have been studied to analyze topological features in data. This has been facilitated by the development of tools for capturing topological features, such as contour trees, merge trees, Reeb graphs, extremum graphs, and persistence diagrams.  A survey paper by Yan \etal \cite{2021-Yan-Scalar-Field-Descriptors} discusses various topological descriptors for the comparison of scalar fields. Various techniques for matching shapes based on topological features have been developed by different groups. Hilaga \etal~\cite{hilaga2001topology} studied the matching of shapes by using a data-structure called multi-resolution Reeb graph. Zhang \etal~\cite{Zhang2004FastMO} proposed a similarity measure based on dual-contour trees and applied it in the classification of protein structures into different categories.  Fabio \etal~\cite{Fabio2016} proposed an edit distance between Reeb graphs and showed its stability. Various distance measures between Reeb graphs such as functional distortion metric \cite{2014-bauer-ReebGraph} and interleaving distance \cite{Silva2016} have been developed which are stable under small perturbations of functions. However, the generalization of these techniques to multi-fields is a challenging task.

Persistent Homology \cite{2002-Edelsbrunner-Persistence,2005-Zomorodian-ComputingPersistentHomology,2009-Chazal-ProximityOfPersistenceModules} provided a topological descriptor for capturing the persistent homological features of high-dimensional datasets based on persistence diagram. Cohen-Steiner \etal~ \cite{2007-Cohen-Steiner-Bottleneck} proposed the bottleneck distance between persistence diagrams and proved it to be stable. Carri\`ere \etal~ \cite{2016-Carriere-ReebGraphs} showed that, locally, the bottleneck distance between Reeb graphs is as discriminative as the functional distortion and interleaving distances. Dey \etal~ \cite{dey2015comparing} developed a stable distance measure between metric graphs based on persistence diagrams. Favelier et al. \cite{2019-Favelier-Persistence-Atlas} proposed a persistence based framework for analyzing clusters and trend variability in ensemble data.

Multi-field topology, being richer than scalar topology, captures more prominent features leading to a better understanding of data. Singh \etal~\cite{2007-Mapper} proposed a \emph{mapper} data-structure to capture the topology of high dimensional point cloud data. Edelsbrunner \etal~ \cite{2008-edels-reebspace} proposed an algorithm for computing the Reeb space. Huettenberger \etal \cite{2013-Huettenberger-Multi-field-Topology} extended the notion of topological structures for multi-fields based on the Pareto optimality and Pareto dominance, and demonstrated their applicability in fluid-flow simulations. Carr \etal~ \cite{JCN_paper} proposed the Joint Contour Net (JCN), which is a quantized approximation of the Reeb space. Duke \etal~\cite{2012-Duke-VisWeek} showed the effectiveness of JCN in visualizing nuclear scission using multivariate density functional theory (DFT) data. Chattopadhyay \etal~  \cite{2014-EuroVis-short} proposed the multi-dimensional Reeb graph (MDRG), which is a hierarchical decomposition of the Reeb space into a set of Reeb graphs in different dimensions and applied this structure for multivariate topology simplification \cite{2015-Chattopadhyay-CGTA-simplification}.
Dey \etal~\cite{2016-MultiscaleMapper} developed the multi-scale mapper (MSM), which is a tower of simplicial complexes connected by simplicial maps, and proved its bottleneck-stability. Agarwal \etal~ \cite{2019-Agarwal-histogram} proposed a distance between multi-fields based on their fiber-component distributions. Recently, Ramamurthi \etal~ \cite{2021-Ramamurthi-MRS} proposed a similarity measure between multi-resolution Reeb spaces, and showed its effectiveness in detecting prominent features in multivariate DFT data. In \secref{sec:experiments}, we compare the proposed method with distances based on fiber-component distributions, MSMs, and MRSs.

The theory of multiparameter persistence \cite{2007-Carlsson-MultiDimensionalPersistence,2010-Carlsson-computing-multidimensional-persistence} deals with the extension of one-parameter persistence to multiparameter based on persistence modules. A survey paper by Botnan \etal \cite{2023-botnan-introduction-multiparameter-persistence} gives a good overview of the recent developments in this direction. Recently, Dey \etal\cite{2022-Dey-Algorithm-Decomposing-Multiparameter-persistence-modules} presented an algorithm for the decomposition of the multiparameter persistence module into indecomposables, which are the counterparts of the bars in the case of one-parameter persistence modules. However, the time complexity for computing the decomposition of a multiparameter persistence module corresponding to a simplicial filtration consisting of $|s|$ simplices, is at least $\cO(|s|^4)$. To address this issue, Loiseaux \etal\cite{2022-Loiseaux-approximation-multiparameter-persistence-modules} introduced a technique that can compute an approximate decomposition for multiparameter persistence modules in lesser time. Some invariants of persistence modules have been developed, such as the Hilbert function, the rank invariant, and multigraded Betti numbers \cite{2023-botnan-introduction-multiparameter-persistence}. Scaramuccia \etal\cite{2020-Scaramuccia-multiparamter-persistent-homology-discrete-morse-theory} introduced a technique for computing multiparameter persistent homology based on \emph{discrete Morse Theory}. Carrière \etal \cite{2020-Carrire-Multiparameter-Persistence-image} developed a descriptor for multiparameter persistence modules by generalizing the \emph{persistence image} of a persistence diagram \cite{2017-Adams-Persistence-Images}.

Lesnick \etal \cite{2015-Lesnick-Interleaving-Distance-Multidimensional-Persistence} generalized the interleaving distance between $1$-parameter persistence modules to multiparameter persistence modules by characterizing \emph{$\epsilon$-interleavings} of multiparameter persistence modules. Dey \etal \cite{2018-Dey-Bottleneck-Distance-Persistence-Modules} developed a polynomial time algorithm for computing the bottleneck distance between interval decomposable persistence modules. Cerri \etal \cite{2013-Cerri-Matching-Distance} introduced the matching distance, which is based on the application of the weighted bottleneck distance on the modules restricted to affine lines. Further, it is shown to serve as a lower bound to the interleaving distance \cite{2014-Landi-Rank-Invariant-Stability}. 
Corbet \etal \cite{2019-Corbet-Multi-parameter-persistent-homology} proposed an inner product based distance measure between multifiltrations. Recently,  generalizations of the Wasserstein distance to multiparameter persistence modules have been developed \cite{2021-Bjerkevik-Wasserstein-distance-multiparameter-persistence}.
Kerber \etal \cite{2019-Kerber-Matching-Distance} developed an algorithm for computing the matching distance between $2$-parameter persistence modules in polynomial time. However, its time complexity is quite high. Hence, an approximation algorithm with improved time complexity has been proposed \cite{2019-Kerber-Approximation-Matching-Distance} based on the findings in \cite{2011-Biasotti-Matching-Distance}. Further, to determine the matching distance between two bifiltrations, the critical values of the simplices in the bifiltrations are projected onto lines in $\R^2$ (called \emph{slices}). However, multiple values may overlap on the same point of a line, leading to a loss of information. 
\begin{figure*}[t]
        \centering
        \includegraphics[width=\textwidth]{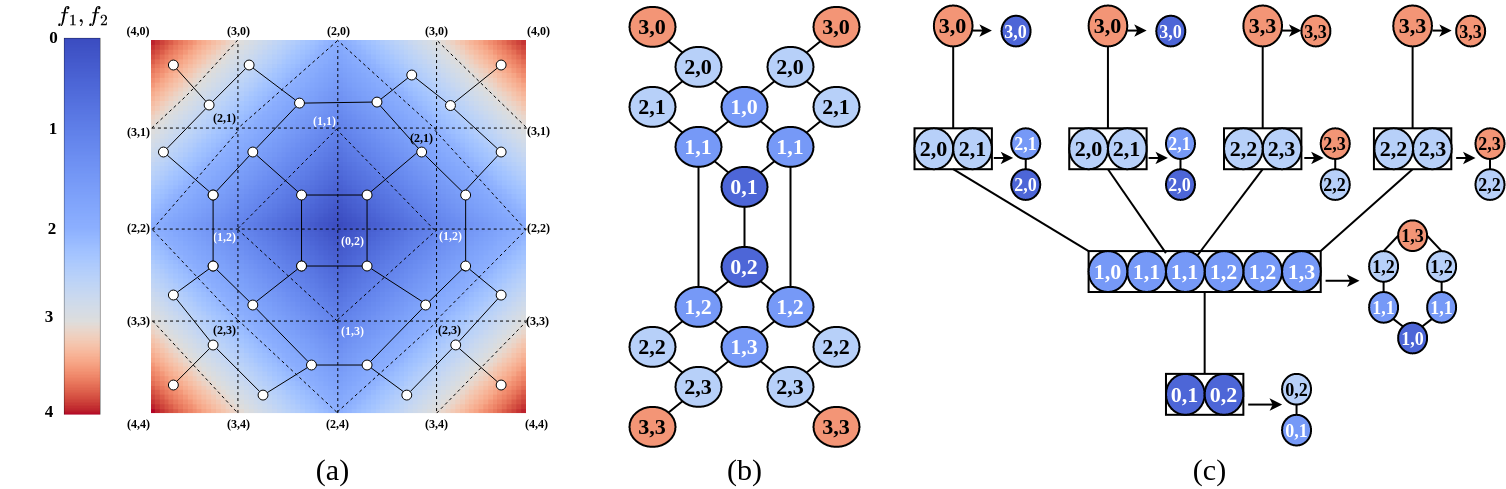}
        \caption{(a) A PL bivariate field applied on a 2D mesh: The mesh  is colored by applying the color map for the first field $(f_1)$. The dashed edges are the boundaries of the joint contours and the adjacency graph of the joint contour fragments  are shown using white nodes and bold edges. (b) JCN at $4 \times 4$ levels of quantization: the coloring of the nodes is based on the values of the first field ($f_1$). Adjacent nodes of the adjacency graph in (a) having the same quantized field values are merged to obtain a single node in the JCN. 
         (c) MDRG constructed using the algorithm in \cite{2015-Chattopadhyay-CGTA-simplification}: the coloring of the nodes in the first (second) dimension are based on the values of $f_1 (f_2)$. Figure reproduced from \cite{2021-Ramamurthi-MRS}.}
        \label{fig:JCN_Mdrg}
\end{figure*}
In this paper, we compare the performance of this approximate matching distance with the proposed distance in shape matching.

In the current paper, our goal is to develop a distance measure between two multi-fields by considering the  persistent features of the corresponding Reeb spaces. 
However, a persistence diagram corresponding to a multifiltration is not known. Moreover, indecomposables of a multiparameter persistence module are complicated and cannot be completely characterized as the bars in one parameter case. Thus, defining a distance between two multi-fields based on indecomposables is challenging \cite{2022-Dey-TDA-book}.
Therefore, in the present work, we study the persistence of a quantized Reeb space or JCN  by a hierarchical decomposition of the JCN into a family of Reeb graphs. The persistence of each Reeb graph can be represented by a persistence diagram using a one-parameter filtration. By combining the persistent features from the component persistence diagrams, we define a multi-dimensional persistence diagram (MDPD) corresponding to a quantized Reeb space. For a JCN of a bivariate field, our algorithm takes $\cO(2|s|^3)$ time for computing the MDPD, where $|s|$ is the total number of vertices and edges in the JCN. We also propose a distance measure between two quantized Reeb spaces based on their MDPDs by extending the Wasserstein distance between two Reeb graphs.

\section{Background}
\label{sec:background}
In this section, we describe the necessary mathematical background to understand our method. A list of notations used in the paper is provided in Appendix \ref{appendix:notations}.

\subsection{Multi-Field}
\label{subsec:multifield}
Let $\cM$ be a compact $m$-manifold. A multi-field or map with $n$ component scalar fields is a continuous map $\f=(f_1,\,f_2,\,\ldots,\,f_n): \cM\rightarrow
\mathbb{R}^n$. Particularly, for the case of $n=1$, $\f$ is a scalar field. For a value $c \in \R^n$, the inverse $\f^{-1}(\mathbf{c})$ is called a \emph{fiber}. Each connected component of a fiber is a \emph{fiber-component} \cite{Saeki2014,2004-Saeki}. Particularly, for the case of a scalar field $f: \cM \rightarrow \R$, the terms fiber and fiber-component are called \emph{level set} and \emph{contour}, respectively.

\textbf{PL Multi-Field:} In scientific visualization most of the data comes as a discrete set of real numbers at the grid points (vertices) of a mesh in a volumetric domain. Let $\M$ be a triangulation of $\cM$ with its vertices containing the data values.  Let $\mathbf{V}(\M)=\{\bv_0, \bv_1, \ldots, \bv_p\}$ be the set of vertices of $\M$. Mathematically, a multi-field data of dimension $n$ can be described as a vertex map $\hat{\f}=(\hat{f_1},\hat{f_2},\ldots,\hat{f_n}):\mathbf{V}(\mathbb{M})\rightarrow \mathbb{R}^n$ such that each vertex is mapped to a $n$-tuple consisting of scalar values. From this map, we obtain a piecewise-linear (PL) multi-field $\f=(f_1,f_2,\ldots,f_n):\mathbb{M}\rightarrow \mathbb{R}^n$ as
$\f(\x)=\sum_{i=0}^p\alpha_i \hat{\f}(\bv_i)$, where each $\x \in \sigma$ (a simplex of $\M$) corresponds to a unique convex combination of its vertices which can be expressed as  $\x=\sum_{i=0}^p\alpha_i \bv_i$ with $\alpha_i\geq 0$ and $\sum_{i=0}^p \alpha_i=1$. We note, $\f$ is continuous and its restriction over the simplices of $\M$ is linear. In the current paper, we consider PL fields with $m \geq n$.

\subsection{Reeb Space}
\label{subsec:reebspace}
For a multi-field $\f: \cM \rightarrow \R^n$ (where $m\geq n$), its Reeb space $\RS_{\f}$ is a quotient space, where points lying in the same fiber-component are considered to be equivalent \cite{2008-edels-reebspace}. In other words, $\RS_{\f}$ is obtained by contracting each fiber-component to a point. In particular, for a scalar field $f: \cM \rightarrow \R$, the Reeb space of $f$ is called a Reeb graph, which is a quotient space corresponding to the equivalence between the points belonging to the contour. The quotient map of $\RS_{\f}$ is given by $q_{\f}: \cM \rightarrow \RS_{\f}$ and the composition of $q_{\f}$ with the unique continuous map $\overline{\f} : \RS_{\f} \rightarrow \R^n$ is called the Stein factorisation of $\f$. The relationship between $\f$, $q_{\f}$, and $\bar{\f}$ is illustrated in the following commutative diagram.
\begin{center}
\begin{tikzcd}[column sep=normal]
\cM \arrow{dr}[swap]{q_\f}\arrow{rr}{\f} & & \mathbb{R}^n\\
& \RS_\f \arrow{ur}[swap]{\bar{\f}} &
\end{tikzcd}
\end{center}

\subsection{Joint Contour Net or Quantized Reeb Space} 
\label{subsec:jcn}
For a triangulation $\M$ of the manifold $\cM$, the Joint Contour Net (JCN) of a PL multi-field $\f: \M \rightarrow \R^n$ is a quantized approximation of the Reeb space $\RS_{\f}$ \cite{JCN_paper}. To construct the JCN, the range of $\f$ is subdivided or quantized into $Q = q_1 \times q_2\times \ldots \times q_n$ quantization levels, where the range of $f_i$ is subdivided into $q_i$ levels. For each quantized range value in $Q$, instead of a fiber, a \emph{quantized fiber} is obtained and each of its connected components is called a \emph{quantized fiber-component} or \emph{joint contour}. The JCN is a graph that captures the adjacency among quantized fiber-components. Each node in the JCN corresponds to a joint contour and an edge between two nodes indicates the adjacency between the corresponding joint contours. \figref{fig:JCN_Mdrg}(b) shows the JCN of a bivariate field. In this paper, we propose a technique for comparing two multi-fields by computing a distance between their respective quantized Reeb spaces or JCNs.

\subsection{Multi-Dimensional Reeb Graph}
\label{subsec:mdrg}
A multi-dimensional Reeb graph (MDRG) is a hierarchical decomposition of a Reeb space proposed by Chattopadhyay \etal~\cite{2015-Chattopadhyay-CGTA-simplification}. Let $\RS_\f$ be the Reeb space of a bivariate field $\f=(f_1,f_2):\cM\rightarrow \mathbb{R}^2$. We can consider a decomposition of this structure as follows. First, consider the Reeb Graph $\RG_{f_1}$ of the field $f_1$. For each $p\in\RG_{f_1}$ consider the restricted field $\widetilde{f_2^p}\equiv f_2|_{C_p}: C_p\rightarrow \mathbb{R}$, where $C_p:=q_{f_1}^{-1}(p)$ is the contour corresponding to $p$. Then consider the Reeb Graph $\RG_{\widetilde{f_2^p}}$ corresponding to each of these restricted scalar fields $\widetilde{f_2^p}$. The hierarchical decomposition of the Reeb Space $\RS_\f$ into the Reeb Graphs $\RG_{f_1}$ and  $\RG_{\widetilde{f_2^p}}$ for each $p\in\RG_{f_1}$ is called the Multi-Dimensional Reeb Graph (MDRG) and is denoted by $\MDRG_\f$. Thus we define the MDRG of $\f=(f_1,f_2)$ as:
\begin{footnotesize}
\begin{align}
\label{eqn:mdrg}
\MDRG_\f &:=\left\{(p_1, p_2): p_1\in \RG_{f_1}, p_2\in \RG_{\widetilde{f_2^{p_1}}}\right\}.
\end{align}
\end{footnotesize}

\noindent
For a general map $\f=(f_1, f_2,\ldots,f_n):\cM \rightarrow \mathbb{R}^n$ (for an $m$-dimensional manifold $\cM$ and $m\geq n\geq 2$) the definition can be generalized as:
\begin{align}
&\MDRG_\f:=\bigg\{(p_1, p_2,\ldots,p_n):p_1\in\RG_{f_1},p_2\in \RG_{\widetilde{f_2^{p_1}}},\nonumber\\
&\hspace{1.7cm}  \,\ldots,\,p_n\in \RG_{\widetilde{f_n^{p_{n-1}}}} \bigg\}.
\end{align}

In  \secref{sec:mdpd}, we design a distance measure between two Reeb spaces based on their MDRGs. 
However, for a continuous map the collection of Reeb graphs in an MDRG is infinite. Therefore, 
for the computational purpose, we consider the quantized Reeb spaces or JCNs to compute our distance measure. \figref{fig:JCN_Mdrg}(c) shows the MDRG corresponding to a quantized Reeb space.

\subsection{Persistence Diagram and Reeb Graph}
\label{subsec:reeb-graph-persistence-diagram}
The persistence diagram is a topological descriptor that captures the topology of a scalar field $f$ through a multiset of points in $\overline{\R}^2$, where $\overline{\R} = \R \cup \infty$. Each point in a persistence diagram of a fixed dimension encodes the birth and death of a homological feature of that dimension.  In particular, connected components are  $0$-dimensional features and loops are $1$-dimensional features. A point $(a,b)$ in a persistence diagram indicates that a homological feature is born at `$a$' and dies at `$b$', and its persistence is given by $b - a$. We refer the reader to Appendix \ref{appendix:persistence-diagrams} for a detailed description of persistence diagrams.

\begin{figure}
    \centering
    \includegraphics[width=0.47\textwidth]{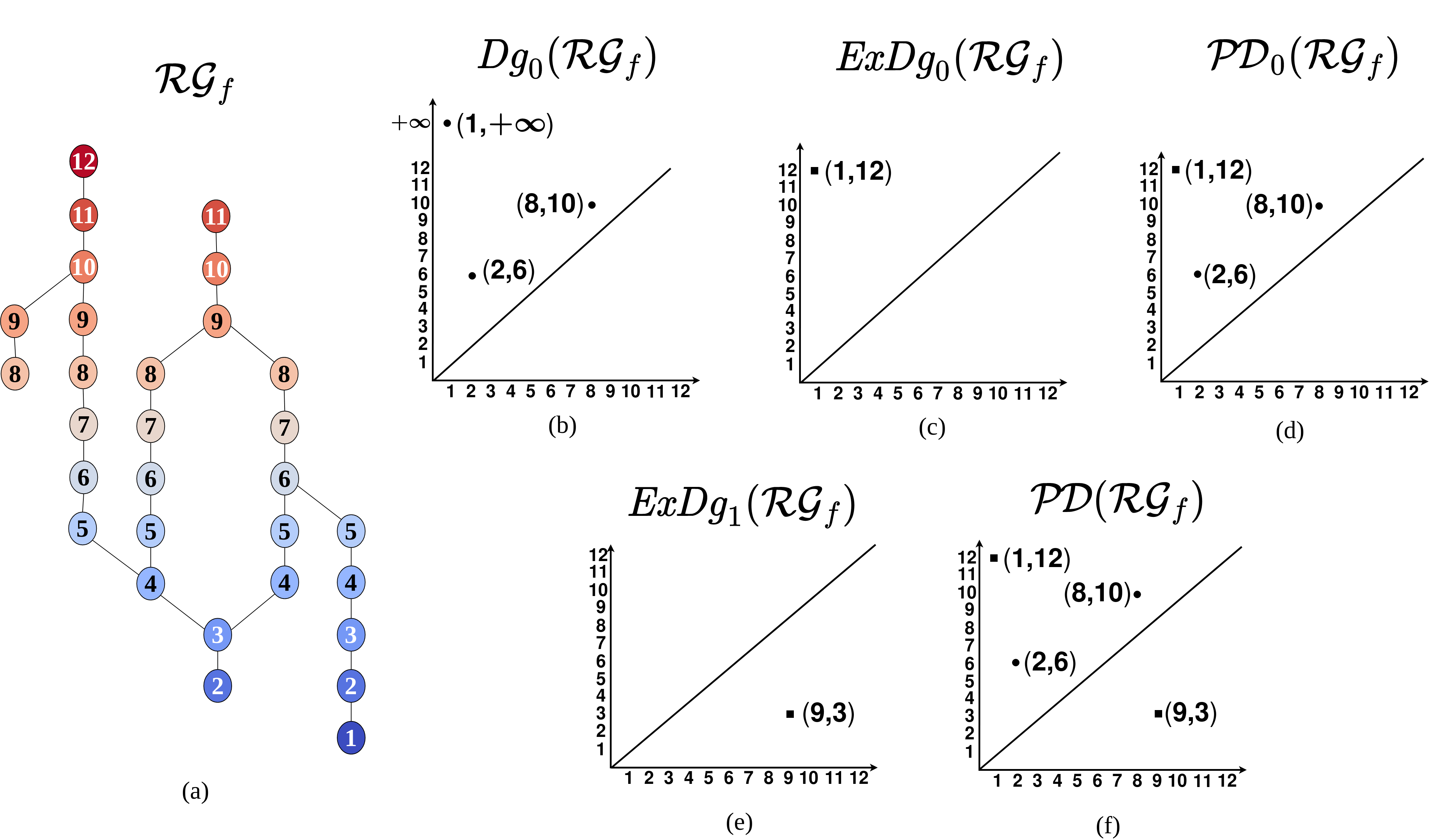}
     \caption{(a) Reeb graph of a real-valued function $f$. (b) $0$-th ordinary persistence diagram $\Dg_0(\RG_f)$. (c) $0$-th extended persistence diagram $\ExDg_0(\RG_f)$. (d) $\PD_0(\RG_f)=\Dg_0(\RG_f)\cup \ExDg_0(\RG_f)\setminus \{(1,\infty)\}$. (e) $1$-st extended persistence diagram $\ExDg_1(\RG_f)$. (f) $\PD(\RG_{f}) = \PD_{0}(\RG_{f}) \cup \ExDg_1(\RG_{f})$. The points in ordinary and extended persistence diagrams are denoted by circular points and squares, respectively.}
    \label{fig:reeb-persistence}
\end{figure}

The persistence diagram corresponding to the Reeb graph of a scalar field is useful in identifying significant features from the Reeb graph. Consider the metaphor of a mountain range, where elevation on earth can be taken as a scalar field \cite{2004-Agarwal-Elevation}. Then each peak corresponds to an arc in the Reeb graph, and a point in the associated persistence diagram. The persistence of a point indicates the prominence of the peak, i.e. the height of its summit above the lowest contour line which encompasses it but doesn't enclose a higher peak. In particular, peaks with higher prominence correspond to points with higher persistence. Further, the persistence diagram provides a compact representation for a Reeb graph, and the computation of distances between Reeb graphs based on their persistence diagrams is more tractable than directly comparing Reeb graphs (see Table $2$ in \cite{2021-Yan-Scalar-Field-Descriptors} for more details).

The persistent homological features of a Reeb graph $\RG_{f}$ can be encoded in $0$- and $1$-dimensional persistence diagrams~\cite{2014-bauer-ReebGraph}. The $0$-th ordinary persistence diagram $\Dg_0(\RG_f)$ encodes the birth and death of sublevel set components in a filtration. The $0$-th extended persistence diagram $\ExDg_0(\RG_{f})$ captures the range of the function $\bar{f}$ and the $1$-st extended persistence diagram $\ExDg_1(\RG_{f})$ captures the loops in $\RG_{f}$. We use the notation $\PD_0(\RG_f)$ to denote the union of $\Dg_0(\RG_f)$ and $\ExDg_0(\RG_f)$, excluding the point with infinite persistence in  $\Dg_0(\RG_f)$. Finally, we obtain a single persistence diagram $\PD(\RG_{f})$ by taking the union of $\PD_0(\RG_f)$  and  $\ExDg_1(\RG_{f})$ which is used for computing the multi-dimensional persistence diagram (that will be discussed in \secref{sec:mdpd}). \figref{fig:reeb-persistence} illustrates the construction of $\PD(\RG_{f})$ from a Reeb graph. 
\begin{figure*}
    \centering    \includegraphics[width=0.83\textwidth]{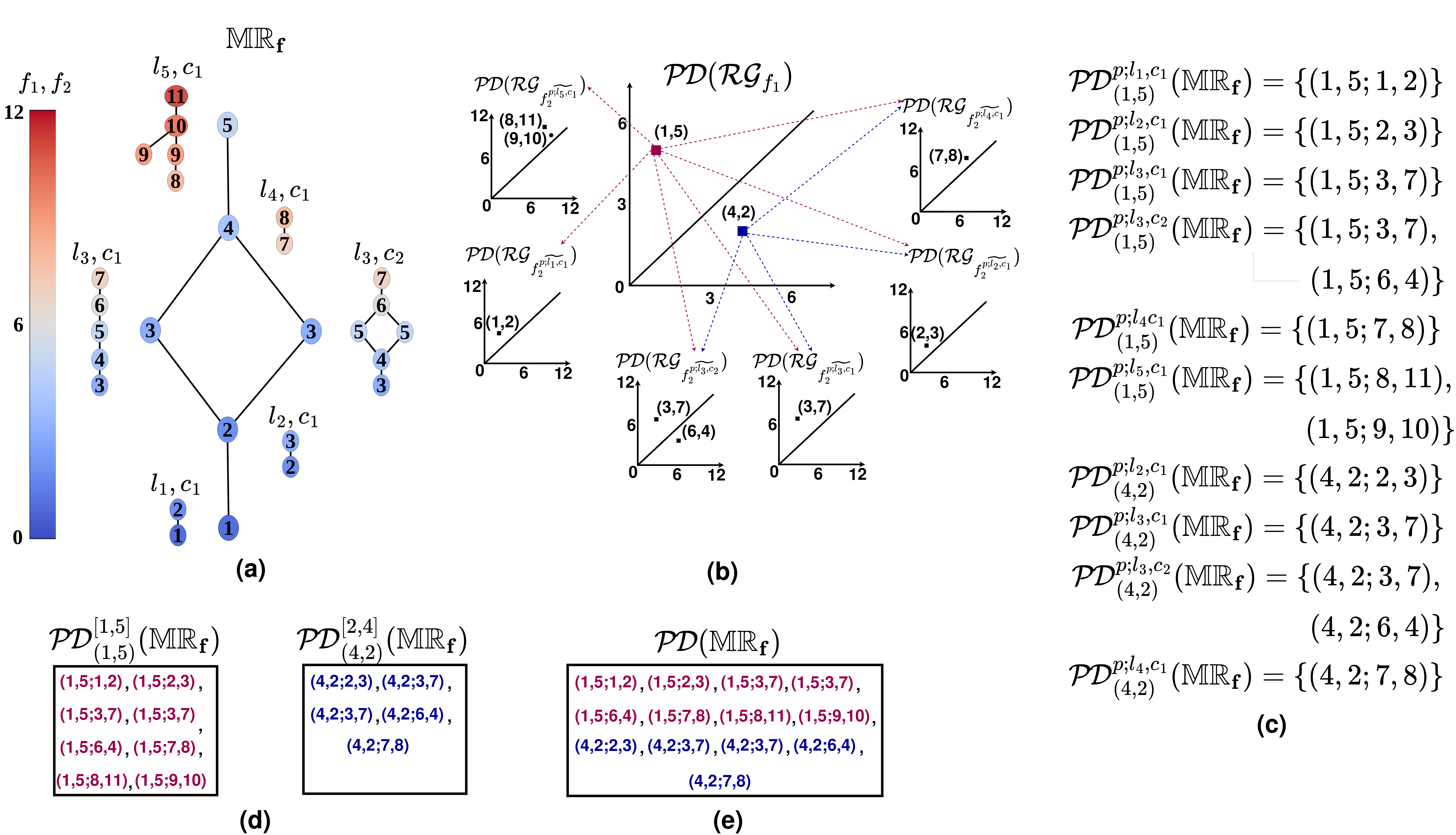}
    \caption{Construction of MDPD from MDRG. (a) MDRG of a bivariate field $\f = (f_1, f_2)$. The nodes in the Reeb graphs of the first (second) dimension are colored by applying the color map for the values of $f_1$ ($f_2$). (b) Persistence diagrams corresponding to the component Reeb graphs of $\MDRG_{\f}$. To denote the level set and contour explicitly,  we also use the notation of the Reeb graph  $\RG_{\tilde{f_2^{p}}}$ by $\RG_{\tilde{f_2^{p;\;l,c}}}$, which corresponds to a contour $c$ of level set $l$ of $f_1$. An arrow from a point $\ux_1$ in $\PD(\RG_{f_1})$ to $\PD(\RG_{\widetilde{f_2^{p;l,c}}})$ indicates that $\bar{f_1}(p;l,c) \in \pI(\ux_1)$ and (c) shows the MDPD of $\f$ relative to $p;l,c$ and $\ux_1$. (d) MDPD of $\f$ relative to each point $\ux_1 \in \PD(\RG_{f_1})$ and persistence interval $\pI(\ux_1)$. (e) Points in the MDPD $\PD(\MDRG_{\f})$. A point $(a,b;c^p,d^p)$ in $\PD(\MDRG_{\f})$ is assigned the same color as the point $(a,b)$ in $\PD(\RG_{f_1})$.
}
    \label{fig:MDPD_Points}
\end{figure*}
We refer the reader to Appendix \ref{appendix:reeb-graph-persistence-diagram} for further details on the construction of the persistence diagrams corresponding to a Reeb graph.

\subsection{Wasserstein Distance between Reeb Graphs}
\label{subsec:wasserstein-distance-reeb-graphs}
Wasserstein distance is commonly known as the earth mover’s distance \cite{2008-villani-book, 2000-Rubner-Earth-Mover-Distance} where the goal is to transform one earth distribution to another by doing the least amount of work in moving piles of earth from the first distribution to the second. Similarly, the idea behind computing the Wasserstein distance between two persistence diagrams is to find the most efficient way of moving the points in the first diagram to be aligned with the points in the second diagram. Let $\Dg_k(\RG_f)$ and $\Dg_k(\RG_g)$ be the  $k$-th ordinary persistence diagrams corresponding to the Reeb graphs $\RG_f$ and $\RG_g$, respectively. Then, similar to the  \emph{Bottleneck distance} between two Reeb graphs by Bauer \etal \cite{2014-bauer-ReebGraph}, we define the \emph{Wasserstein distance} between $\RG_f$ and $\RG_g$ as: 
\begin{footnotesize}
\begin{equation}
\begin{aligned}
\label{eqn:wasserstein-distance-reeb-graph}
&d_{W,k,q}(\RG_f, \RG_g)\\
&:=\left [ \displaystyle\inf_{\eta: \Dg_k(\RG_f)\rightarrow
    \Dg_k(\RG_g)}\sum_{x\in \Dg_k(\RG_f)}\|x-\eta(x)\|_{\infty}^q\right ]^{1/q}
\end{aligned}
\end{equation}
\end{footnotesize}
\noindent    
for any positive real number $q$ and for each dimension $k$. Here $\eta: \Dg_k(\RG_f)\rightarrow \Dg_k(\RG_g)$ is any bijection between $\Dg_k(\RG_f)$ and $\Dg_k(\RG_g)$ and $\|.\|_{\infty}$ denotes the $l_\infty$ norm between two points, i.e., for two points  $\x=(x_1,x_2)$ and $\y=(y_1,y_2)$ in $\mathbb{R}^2$, we have $\|\x-\y\|_{\infty}=\max\{|x_1-y_1|, |x_2-y_2|\}$. A similar definition also holds between extended persistence diagrams corresponding to the Reeb graphs. The Wasserstein distance (\eqnref{eqn:wasserstein-distance-reeb-graph}) between two Reeb graphs satisfies the following property.

\begin{lemma}
$d_{W,k,q}(\RG_f, \RG_g)$ is a pseudo-metric.
\end{lemma}

In the current paper, we propose a distance to extend the above Wasserstein distance between two Reeb graphs to quantized Reeb spaces. To develop our distance, first, we decompose each quantized Reeb space into its corresponding
MDRG. Based on this, we compute a multi-dimensional persistence diagram representation for each  quantized Reeb space as described in the next section.

\section{An Algorithm for Computing a Distance Measure between Quantized Reeb Spaces} 
\label{sec:persistent-mdrg}
In this section, we first compute the MDRG structure corresponding to each JCN as described by Chattopadhyay \etal~\cite{2014-EuroVis-short, 2015-Chattopadhyay-CGTA-simplification}.
Then we define the notion of a multi-dimensional persistence diagram (MDPD) to encode the persistent features of the component Reeb graphs of the MDRG. This representation is used later to design a topological distance between two quantized Reeb spaces. For simplicity, we describe our algorithm for a bivariate field $\f=(f_1,f_2)$. The extension of the method for $n$ number of fields is provided in Appendix \ref{appendix:generalization-MDPD-n-fields}.

\subsection{Multi-Dimensional Persistence Diagram}
\label{sec:mdpd}
Analyzing persistent features in data across multiple parameters can yield deeper insights into it. For instance, consider investigating temperature patterns in mountain ranges. The temperature at a particular point depends both on its latitude and elevation (altitude). The latitude influences the angle of sunlight it receives. Locations near the equator receive sunlight more directly, resulting in higher temperatures. Conversely, the polar regions receive sunlight at a more oblique angle, which generally leads to cooler temperatures. Similarly, as altitude increases, temperature typically decreases. Consequently, mountaintops are generally cooler than their bases. The persistence diagram corresponding to the Reeb graph of the elevation field captures peaks along with their prominence. A point $(a,b)$ in the persistence diagram represents a peak with 
base altitude $a$ and summit altitude $b$. 
Within the altitudes spanning between $a$ and $b$, the variation of latitudes along various peaks can be captured by persistence diagrams (of the Reeb graphs) corresponding to the latitudes.  The persistent features of both the altitude and latitude can jointly be represented in an MDPD. This consolidated representation can then be utilized in examining how temperature depends on both of these parameters.

Consider the MDRG  of the bivariate field $\f=(f_1,f_2)$, denoted by $\MDRG_\f$, as defined in \eqnref{eqn:mdrg}. $\MDRG_\f$ consists of Reeb graphs $\RG_{f_1}$ and $\RG_{\tilde{f_2^{p}}}$ for $p\in \RG_{f_1}$. We obtain the persistence diagram $\PD(\RG_{f_1})$ corresponding to  $\RG_{f_1}$  
and the persistence diagram $\PD(\RG_{\tilde{f_2^{p}}})$  corresponding to $\RG_{\tilde{f_2^{p}}}$ for each $p\in \RG_{f_1}$, as described in \secref{subsec:reeb-graph-persistence-diagram}. For a birth-death point $\ux_1=(a_1,b_1)\in \PD(\RG_{f_1})$, its \emph{persistence interval} is defined as $\pI(\ux_1)=[\min\{a_1, b_1\}, \max\{a_1,b_1\}]$. We note, $\PD(\RG_{f_1})$ consists of both $0$- and $1$-dimensional persistent features.
    For $0$-dimensional persistent features, birth is less than death, whereas, birth is greater than death for $1$-dimensional persistent features. Intuitively, an MDPD corresponding to an MDRG of a bivariate field, captures the persistent features of its Reeb graphs in various dimensions, jointly.
More explicitly, with each persistent feature $\ux_1$ in the persistence diagram corresponding to the Reeb graph in the first dimension, the MDPD associates the persistent features of the second-dimensional Reeb graphs that occur within the life-span of $\ux_1$.

Mathematically, we define the MDPD of $\MDRG_\f$ using the following steps.
First, the MDPD of $\MDRG_\f$ relative to a birth-death point $\ux_1\in \PD(\RG_{f_1})$ and a node $p\in \RG_{f_1}$, is defined as:
\begin{footnotesize}
\begin{eqnarray}
   \label{eqn:MDPD-1}
    \PD^{p}_{\ux_1}(\MDRG_\f):=&\{\ux_1\}\times \PD(\RG_{\tilde{f_2^{p}}})\nonumber\\
    =&\{(\ux_1;\,\ux_2^p):\ux_2^p\in \PD(\RG_{\tilde{f_2^{p}}})\}.
\end{eqnarray}
\end{footnotesize}
\noindent
In the second step, we consider the second-dimensional persistence diagrams corresponding to the nodes $p\in \RG_{f_1}$ such that $\bar{f}_1(p)\in \pI(\ux_1)$. Therefore, we define MDPD of $\MDRG_\f$ relative to a birth-death point $\ux_1\in \PD(\RG_{f_1})$ and persistence interval $\pI({\ux_1})$ as:
\begin{footnotesize}
\begin{eqnarray}
   \label{eqn:MDPD-2}
\PD^{\pI(\ux_1)}_{\ux_1}(\MDRG_\f):=\bigcup_{\{p\in \RG_{f_1}: \bar{f}_1(p)\in \pI(\ux_1)\}}\PD^{p}_{\ux_1}(\MDRG_\f).
\end{eqnarray}
\end{footnotesize}
\noindent
Finally, by considering  $\PD^{\pI(\ux_1)}_{\ux_1}(\MDRG_\f)$  for all birth-death points $\ux_1\in \PD(\RG_{f_1})$, we define the MDPD of $\MDRG_\f$  as:
\begin{footnotesize}
\begin{eqnarray}
    \label{eqn:MDPD}
    \PD(\MDRG_\f):=\bigcup_{\ux_1\in\PD(\RG_{f_1})}\PD^{\pI(\ux_1)}_{\ux_1}(\MDRG_\f).
\end{eqnarray}
\end{footnotesize}
Thus, from equations (\ref{eqn:MDPD-1}), (\ref{eqn:MDPD-2}), and (\ref{eqn:MDPD}), we  have:\\
\begin{footnotesize}
\begin{eqnarray}
\PD(\MDRG_\f):=
   \displaystyle \bigcup_{\ux_1\in\PD(\RG_{f_1})}\bigcup_{\{p\in \RG_{f_1}: \bar{f}_1(p)\in \pI(\ux_1)\}}\{(\ux_1;\,\ux_2^p):\nonumber\\
   \ux_2^p\in \PD(\RG_{\tilde{f_2^{p}}})\}
   \label{eqn:MDPD-2-fields-full-version}
\end{eqnarray}
\end{footnotesize}
which is a multiset of points in $\R^2 \times \R^2$ or $\R^4$. \figref{fig:MDPD_Points} shows an example of MDPD for a bivariate field. Algorithm \ref{algo:MDPD} gives the pseudo-code for the construction of MDPD from a quantized MDRG.

We note, the \emph{indecomposables} obtained from a multiparameter persistence module  cannot be completely characterized as the bars corresponding to a $1$-parameter persistence module \cite{2022-Dey-TDA-book}. However, the persistent features in an MDPD of a bivariate field can be characterized as rectangles in $\R^2$. In particular, a point $(\ux_1;\ux_2^{p}) \in\PD(\MDRG_{\f})$ represents the persistence of a pair of homology classes $(\gamma_1,\gamma_2^p)$ corresponding to the filtrations of $\RG_{f_1}$ and $\RG_{\tilde{f_2^{p}}}$, respectively. The  persistence measure of $(\gamma_1,\gamma_2^{p})$ is considered as the area of the rectangle obtained by the cartesian product of the persistence intervals $\pI(\ux_1)$ and $\pI(\ux_2^p)$.  In general, the persistence measure of a point in an MDPD corresponding to an $n$-dimensional multi-field is given by the volume of the corresponding $n$-dimensional orthotope (see Appendix \ref{appendix:generalization-MDPD-n-fields} for more details).

\begin{algorithm}
\caption{\sc{ConstructMDPD}}
\label{algo:MDPD}
\textbf{Input:} $\MDRG_{\f}$\\
\textbf{Output:} $\PD(\MDRG_{\f})$
\begin{algorithmic}[1]
\STATE \% \textit{Computing the persistence diagrams of Reeb graphs in $\MDRG_{\f}$}
\STATE Compute $\PD(\RG_{f_1})$
\FOR{$p \in \RG_{f_1}$}
\STATE Compute $\PD \left(\RG_{\widetilde{f_2^p}}\right)$
\ENDFOR
\STATE \%\textit{Constructing the MDPD}
\STATE $\PD(\MDRG_{\f}) = \{\}$
\FOR{$(a_1,b_1) \in \PD(\RG_{f_1})$}
\FOR{$p \in \RG_{f_1} \text{ with }\bar{f_1}(p) \in \pI((a_1,b_1))$}
\STATE $S = \left\{(a_1,b_1;a_2^p,b_2^p) | (a_2^p,b_2^p) \in \PD\left(\RG_{\widetilde{f_2^{p}}}\right)\right\}$
\STATE $\PD(\MDRG_{\f}) = \PD(\MDRG_{\f}) \cup S$
\ENDFOR
\ENDFOR
\RETURN{$\PD(\MDRG_{\f})$}
\end{algorithmic}
\end{algorithm}

Next, we propose a distance measure between two quantized Reeb Spaces  
based on their MDPDs.

\subsection{Proposed Distance based on MDPDs}
\label{subsec:distance-between-MDRGs}
 Let $\JCN_\f$ and $\JCN_\g$ be the joint contour nets corresponding to the bivariate fields $\f$ and $\g$, respectively, constructed using identical quantization levels. Let,  $\MDRG_{\f}$, $\MDRG_{\g}$ be the corresponding MDRGs. From these MDRGs, we compute the  MDPDs $\PD(\MDRG_\f)$ and $\PD(\MDRG_\g)$ using Algorithm~\ref{algo:MDPD}.
Then we define the \emph{Wasserstein distance} between $\MDRG_{\f}$ and $\MDRG_{\g}$, for any positive real number $q$, as:
\begin{footnotesize}
\begin{equation}
\label{eqn:Wasserstein-Distance-MDPDs}
\begin{aligned}
&d_{W,q}(\MDRG_\f, \MDRG_\g)\\
&:=\displaystyle\left(\inf_{\eta: \PD(\MDRG_\f)\rightarrow
    \PD(\MDRG_\g)}\sum_{x \in \PD(\MDRG_{\f})} \|\x-\eta(\x)\|_{\infty}^q\right)^{1/q}.
\end{aligned}
\end{equation}
\end{footnotesize}
\noindent
Here, $\|.\|_{\infty}$ denotes the $l_\infty$ norm and $\eta: \PD(\MDRG_\f)\rightarrow \PD(\MDRG_\g)$ ranges over all bijections between $\PD(\MDRG_\f)$ and $\PD(\MDRG_\g)$ satisfying the criteria (C1)-(C3), as described below. For the construction of optimal $\eta$, we apply the technique in \emph{Hungarian algorithm} \cite{1955-Kuhn-Hungarian-Algorithm}.  To make the cardinalities of the MDPDs equal, we consider additional diagonal points, obtained by taking the product of diagonal points in the component  persistence diagrams corresponding to each MDPD. 
The bijections $\eta : \PD(\MDRG_{\f}) \rightarrow \PD(\MDRG_{\g})$ satisfy the following criteria.

\begin{enumerate}[label=(C{{\arabic*}})]
    \item \textbf{Matching points from the same levels of the first component fields:}  Let $\pointa = (a_1,b_1;a_2^p,b_2^p) \in \PD(\MDRG_{\f})$. $\eta$ maps $\pointa$ either to a point $\pointb=(c_1,d_1;c_2^{p'},d_2^{p'}) \in \PD(\MDRG_{\g})$ such that $\bar{f_1}(p) = \bar{g_1}(p')$, or $\eta(\pointa)$ is a diagonal point of $\PD(\MDRG_{\g})$. This condition ensures that points in the MDPDs are matched by $\eta$ only when they correspond to the same levels of $f_1$ and $g_1$.
    
    \item \textbf{Topological consistency:} If  $\pointa = (a_1,b_1;a_2^p,b_2^p) \in \PD(\MDRG_{\f})$ is matched with  $\pointb = (c_1,d_1;c_2^{p'},d_2^{p'}) \in \PD(\MDRG_{\g})$ by $\eta$, then for all other points of the form $\x=(a_{i1}, b_{i1}; a_{i2}^{p}, b_{i2}^{p})\in \PD(\MDRG_{\f})$, $\eta(\x)$  will be of the form $(c_{j1}, d_{j1}; c_{j2}^{p'}, d_{j2}^{p'})\in \PD(\MDRG_{\g})$, or a diagonal point of $\PD(\MDRG_{\g})$. That is, all the points in $\PD(\MDRG_{\f})$  obtained from a persistence diagram $\PD(\RG_{\widetilde{f_2^{p}}})$ are mapped to points in $\PD(\MDRG_{\g})$ obtained from the persistence diagram $\PD(\RG_{\widetilde{g_2^{p'}}})$. In other words, no two points in $\PD(\MDRG_{\f})$ coming from $\PD(\RG_{\widetilde{f_2^{p}}})$ are mapped to points in $\PD(\MDRG_{\g})$ coming from two different persistence diagrams $\PD(\RG_{\widetilde{g_2^{p'}}})$ and $\PD(\RG_{\widetilde{g_2^{p''}}})$ with $p' \neq p''$. This condition ensures the topological consistency by mapping all points in $\PD(\MDRG_{\f})$ corresponding to a contour of $f_1$ to points in $\PD(\MDRG_{\g})$ corresponding to a contour of $g_1$.

\item \textbf{Dimension consistency:} $\pointa=(a_1,b_1;a_2^p,b_2^p) \in \PD(\MDRG_{\f})$ is matched with $\pointb=(c_1,d_1;c_2^{p'},d_2^{p'}) \in \PD(\MDRG_{\g})$ by $\eta$ if (i) the dimensions of the persistent homology classes corresponding to $(a_1,b_1)$ and $(c_1,d_1)$ are the same and (ii) the dimensions of the persistent homology classes corresponding to $(a_2^p, b_2^p)$ and $(c_2^{p'}, d_2^{p'})$ are the same.
\end{enumerate}
\noindent
\algoref{algo:Distance} gives an outline of computing the proposed  measure.

\begin{algorithm}
\caption{\sc{ComputeDistance}}
\label{algo:Distance}
\textbf{Input:} $\MDRG_{\f},\; \MDRG_{\g},\; q$\\
\textbf{Output:} $d_{W,q}(\MDRG_{\f}, \MDRG_{\g})$
\begin{algorithmic}[1]
\STATE \% \textit{Compute the MDPDs corresponding to the MDRGs}
\STATE $\PD(\MDRG_{\f})$={\sc ConstructMDPD}$(\MDRG_{\f})$
\STATE $\PD(\MDRG_{\g})$={\sc ConstructMDPD}$(\MDRG_{\g})$
\STATE Compute optimal bijection $\eta$ 
between points in $\PD(\MDRG_{\f})$ and $\PD(\MDRG_{\g})$, satisfying (C1)-(C3), using Hungarian algorithm.
\STATE $d_{W,q} = \left(\sum_{\x \in \PD(\MDRG_{\f})} \|\x-\eta(\x)\|_{\infty}^q\right)^{1/q}$ 
\RETURN{$d_{W,q}$}
\end{algorithmic}
\end{algorithm}

\noindent
\textbf{Details of Line 4 of Algorithm \ref{algo:Distance}.} To understand the computation of line 4 of Algorithm \ref{algo:Distance}, for each fixed $p\in\RG_{f_1}$, we define the subset $\PD^{p}(\MDRG_{\f})$ of $\PD(\MDRG_{\f})$ as:
\begin{footnotesize}
\begin{equation}
    \PD^p(\MDRG_{\f}) := \bigcup_{\ux_1 \in \PD(\RG_{f_1}), \bar{f_1}(p) \in \pI(\ux_1)} \PD_{\ux_1}^{p}(\MDRG_{\f}).
    \label{eqn:PD^p}
\end{equation}
\end{footnotesize}
Thus we have  $\PD(\MDRG_{\f}) = \bigcup_{p \in \RG_{f_1}}\PD^{p}(\MDRG_{\f})$. Similarly, we define $\PD^{p}(\MDRG_{\g})$  for each $p \in \RG_{g_1}$.

We note, the condition (C2) reduces the task of computing bijections between $\PD(\MDRG_{\f})$ and $\PD(\MDRG_{\g})$ to computing bijections between the elements of $\cS_{\f} = \{\PD^{p}(\MDRG_{\f})
| p \in \RG_{f_1}\}$ and $\cS_{\g} = \{\PD^{p'}(\MDRG_{\g})
 | p' \in \RG_{g_1}\}$. For $\PD^{p}(\MDRG_{\f})
 \in \cS_{\f}$ and $\PD^{p'}(\MDRG_{\g})
 \in \cS_{\g}$ with $\bar{f_1}(p) = \bar{g_1}(p')$, we compute:
 \begin{footnotesize}
\begin{equation}
\begin{aligned}
&d(\PD^{p}(\MDRG_{\f}), \PD^{p'}(\MDRG_{\g}))\\
&= \inf_{\eta' :\PD^{p}(\MDRG_{\f})\rightarrow \PD^{p'}(\MDRG_{\g})} \sum_{x\in\PD^{p}(\MDRG_{\f})}\|x - \eta'(x) \|^q
\end{aligned}
\label{eqn:d}
\end{equation}
\end{footnotesize}
where, $\eta'$ ranges over bijections between $\PD^{p}(\MDRG_{\f})$ and $\PD^{p'}(\MDRG_{\g})$ satisfying (C3), computed using Hungarian algorithm \cite{1955-Kuhn-Hungarian-Algorithm}. Finally, $d_{W,q}(\MDRG_{\f}, \MDRG_{\g})$ is computed as:
\begin{footnotesize}
\begin{equation}
\label{eqn:Wasserstein-intermediate}
\begin{aligned}
&d_{W,q}(\MDRG_{\f}, \MDRG_{\g})\\
&=\displaystyle\inf_{\eta'' : \RG_{f_1} \rightarrow \RG_{g_1}} \left(\sum_{p\in \RG_{f_1}} d(\PD^{p}(\MDRG_{\f}), \PD^{\eta''(p)}(\MDRG_{\g}))\right)^{\frac{1}{q}}
\end{aligned}
\end{equation}
\end{footnotesize}
where, $\eta''$ ranges over all maximal bipartite matchings between the nodes of $\RG_{f_1}$ and $\RG_{g_1}$.  
A node $p$ in $\RG_{f_1}$ will be matched to a node $p'$ in $\RG_{g_1}$ by $\eta''$ only if $\bar{f_1}(p) = \bar{g_1}(p')$.
We note, all nodes in $\RG_{f_1}$ may not match with all the nodes of  $\RG_{g_1}$ by $\eta''$, so we extend $\eta''$ as follows. For each unmatched node $p$ in $\RG_{f_1}$, we consider the set of diagonal points $D^{p} = \{(\frac{a_1+b_1}{2}, \frac{a_1+b_1}{2}; \frac{a_2^p+b_2^p}{2}, \frac{a_2^p+b_2^p}{2}) | (a_1,b_1;a_2^p,b_2^p) \in \PD^p(\MDRG_{\f})\}$ and compute $d(\PD^{p}(\MDRG_{\f}), D^p)$. 
Similarly, for every unmatched node $p' \in \RG_{g_1}$, we compute $d(D^{p'},\PD^{p'}(\MDRG_{\g}))$. We construct optimal $\eta''$ using Hungarian algorithm  by computing a cost matrix \cite{1955-Kuhn-Hungarian-Algorithm}.

\subsection{Properties of the Proposed Distance Measure} In this sub-section, we discuss the properties of the Wasserstein distance between MDRGs. 

\noindent
\textbf{Pseudo-metric.}  Let $\f=(f_1,f_2), \g=(g_1, g_2)$ be two bivariate fields. First, we show that the Wasserstein distance between $\MDRG_{\f}$ and $\MDRG_{\g}$ is a pseudo-metric. 

\begin{theorem}
$d_{W,q}(\MDRG_{\f}, \MDRG_{\g})$ is a pseudo-metric.
\label{theorem:pseudo-metric}
\end{theorem}
\begin{proof}
See Appendix \ref{appendix:pseudo-metric}.
\end{proof}

\noindent
\textbf{Stability.} To prove the stability of the MDRG of a bivariate field w.r.t. the proposed distance measure, we need to show that a small perturbation in the bivariate field produces a small variation in the MDRG. To achieve this, we derive an upper bound on the Wasserstein distance between $\MDRG_{\f}$ and $\MDRG_{\g}$ based on (i) the number of vertices in  $\RG_{f_1}$ and $\RG_{g_1}$, (ii) the number of critical points  of $f_1, g_1, \widetilde{f_2^p}$, and $\widetilde{g_2^{p'}}$ where $p\in \RG_{f_1}$ and $p' \in \RG_{g_1}$, and (iii) the  amplitudes of $f_1, f_2, g_1$, and $g_2$. Here, the amplitude of a function $f: \cM \rightarrow \R$ is defined as follows: $Amp(f) = \max_{\x \in \cM} f(\x) - \min_{\x \in \cM} f(\x)$.

\begin{lemma}
\label{lemma:stability}
Let $\f = (f_1, f_2)$ and $\g = (g_1, g_2)$ be bivariate fields defined on a compact $m$-manifold $\cM$ with $m\geq 2$. Then
\begin{footnotesize}
\begin{equation}
   \begin{aligned}
  &d_{W,q}(\MDRG_{\f}, \MDRG_{\g})\\
&\leq\biggl( N_{\RG_{f_1}} C_{f_1} \max_{p \in \RG_{f_1}} C_{\widetilde{f_2^p}} \Bigl(\max\{Amp(f_1), Amp(f_2),\\
&\hspace{4.5cm}Amp(g_1),Amp(g_2)\}\Bigr)^q\\
&\hspace{0.2cm}+N_{\RG_{g_1}} C_{g_1} \max_{p \in \RG_{g_1}} C_{\widetilde{g_2^p}} \Bigl(\max\{Amp(f_1),Amp(f_2),\\
&\hspace{4.5cm}Amp(g_1),Amp(g_2)\}\Bigr)^q\biggr)^{\frac{1}{q}}
   \end{aligned} 
\end{equation}
\end{footnotesize}
where, for a function $f$, $C_{f}$ denotes the number of critical points, $N_{\RG_{f}}$ is the number of vertices in the quantized approximation of its Reeb graph, and $Amp(f)$ denotes the amplitude of $f$.
\end{lemma}
\begin{proof}
See Appendix \ref{appendix:stability}.
\end{proof}

\subsection{Complexity Analysis}
In this sub-section, we analyze the time complexities of constructing the MDPD of a bivariate field from its MDRG (Algorithm \ref{algo:MDPD}) and computing the distance between two MDRGs using MDPDs (Algorithm \ref{algo:Distance}).

\subsubsection{Algorithm \ref{algo:MDPD}: Constructing the MDPD}
\label{subsubsec:complexity-mdpd-construction}
Let $\f= (f_1, f_2)$ be a bivariate field defined on a compact $m$-manifold. First, note that the MDRG of $\f$ is constructed from its JCN using the algorithm in \cite{2015-Chattopadhyay-CGTA-simplification}. 
The time complexity for computing $\MDRG_{\f}$ from $\JCN_{\f}$ is $\cO( 2|V|(|V| + |E| \alpha (|V|) + |V| \log (|V|)))$, where $|V|$ and $|E|$ are the number of vertices and edges, respectively, in $\JCN_{\f}$ \cite{2021-Ramamurthi-MRS}. 
Next, we analyze the time complexity of computing the MDPD from $\MDRG_{\f}$. 

To compute the MDPD, we need to compute the persistence diagrams of the component Reeb graphs in $\MDRG_{\f}$ (lines: $1$-$5$, in Algorithm \ref{algo:MDPD}). The number of vertices and edges in $\RG_{f_1}$ are at most $|V|$ and $|E|$, respectively. Therefore, the number of simplices (total number of vertices and edges) in $\RG_{f_1}$ is at most $|V| + |E|$. The time complexity for the construction of $\PD(\RG_{f_1})$ is $\cO\left((|V| + |E|)^3\right)$ (lines $1$-$2$, Algorithm \ref{algo:MDPD}), as discussed in \cite{book-herbert-computational-topology}. Each node (similarly, edge) in a Reeb graph $\RG_{\widetilde{f_2^p}}$  ($p\in\RG_{f_1}$) of $\MDRG_{\f}$ corresponds to a unique node (edge) in $\JCN_{\f}$. In other words, the total number of nodes (similarly, edges)  in all the Reeb graphs $\RG_{\widetilde{f_2^p}}$ (for all $\; p\in\RG_{f_1}$) is at most the total number of nodes (edges) of the JCN. Thus the total time complexity for computing the persistence diagrams  $\left\{\PD\left(\RG_{\widetilde{f_2^p}}\right) \mid p\in\RG_{f_1}\right\}$ is $\cO((|V| + |E|)^3)$ (lines $3$-$5$, Algorithm \ref{algo:MDPD}).

Next, we see the time complexity of constructing the MDPD (lines $6$-$13$, Algorithm \ref{algo:MDPD}). Each point in the persistence diagram of a Reeb graph corresponds to a pair of critical nodes in the Reeb graph (see \secref{subsec:reeb-graph-persistence-diagram} for more details). Hence, the number of points in $\PD(\RG_{f_1})
$ is at most of $\cO(|V|)$. Similarly, since there is a one-to-one correspondence between vertices of $\JCN_{\f}$ and the nodes of $\RG_{\widetilde{f_2^p}}$, the total number of points in all the persistence diagrams $\left\{\PD\left(\RG_{\widetilde{f_2^p}}\right) \mid p\in\RG_{f_1}\right\}$ is at most of $\cO(|V|)$. Therefore, the time complexity for lines $6$-$13$ of Algorithm \ref{algo:MDPD}  is $\cO(|V|^2)$. Finally, the total time complexity for computing the MDPD from $\JCN_\f$ is $\cO(2|E||V|\alpha(|V|) + 2|V|^2 \log(|V|) + 2(|V| + |E|)^3 + |V|^2)$,  which is equivalent to   $\cO(2(|V|+|E|)^3)$.

\subsubsection{Algorithm \ref{algo:Distance}: Computing Distance between MDRGs}
\label{subsubsec:complexity-distance}
Next, we discuss the time complexity for Algorithm \ref{algo:Distance}. Let $\PD(\MDRG_{\f})$ and $\PD(\MDRG_{\g})$ be the MDPDs of two bivariate fields $\f = (f_1, f_2)$ and $\g = (g_1, g_2)$, respectively. Let $|V_1|$ and $|V_2|$ be the number of vertices in $\JCN_{\f}$ and $\JCN_{\g}$, respectively. We note, the number of points in $\PD(\RG_{f_1})$ is bounded by the number of critical points of $\bar{f_1}$, which is at most the number of critical points $C_{f_1}$  of $f_1$. Since the total number of points in the persistence diagrams $\left\{\PD\left(\RG_{\widetilde{f_2^p}}\right) \mid p\in\RG_{f_1}\right\}$ is bounded by $|V_1|$, the MDPD $\PD(\MDRG_{\f})$ has at most $C_{f_1} |V_1|$ points. Similarly, the number of points in $\PD(\MDRG_{\g})$ is at most $C_{g_1} |V_2|$, where $C_{g_1}$ is the number of critical points of $g_1$. For computing $d_{W,q}(\MDRG_{\f}, \MDRG_{\g})$, in line $4$ of Algorithm \ref{algo:Distance}, we consider bijections between $\PD(\MDRG_{\f})$ and $\PD(\MDRG_{\g})$ satisfying the criteria (C1) - (C3) mentioned in \secref{subsec:distance-between-MDRGs}. However, in the worst case, when we consider all possible bijections, the time complexity for computing an optimal bijection using the Hungarian algorithm \cite{1955-Kuhn-Hungarian-Algorithm} is $\cO((C_{f_1}|V_1| + C_{g_1}|V_2|)^3)$. Therefore, the total time complexity of Algorithm \ref{algo:Distance} is 
$\cO(2(|V_1|+|E_1|)^3)+ \cO(2(|V_2|+|E_2|)^3)+\cO((C_{f_1}|V_1| + C_{g_1}|V_2|)^3)$.

\section{Experimental Results} 
\label{sec:experiments}
In this section, the application of the proposed distance on SHREC $2010$ and Pt-CO bond formation datasets are demonstrated.  We note, to compute the MDPD corresponding to an MDRG, we need to handle the degenerate critical nodes in the component Reeb graphs of the MDRG which is discussed in the next subsection.

\subsection{Handling Degeneracy }
To construct the persistence diagram of the Reeb graph $\RG_{f}$, we require $\RG_{f}$ to be the Reeb graph of a \emph{Morse} function $f$ \cite{2019-Tu-PropogateAndPair}. A real-valued function $f$ is  Morse if it satisfies the following criteria: (i) all the critical points of $f$ are non-degenerate and (ii) no two critical points have the same critical value. To construct the MDPD of an MDRG, we require these criteria to be satisfied by all the Reeb graphs in the MDRG.

A node in a Reeb graph is said to be degenerate if it is of one of the following types: (i) degenerate maximum (up-degree = $0$, down-degree $ > 1$), (ii) degenerate minimum (up-degree $ > 1$, down-degree $ = 0$, (iii) double fork (up-degree $ = 2$, down-degree $= 2$),  and (iv) complex fork (up-degree $>2$, down-degree $> 2$). We remove degenerate nodes by breaking them into non-degenerate critical nodes \cite{2019-Tu-PropogateAndPair}. After removing degeneracy, we need to ensure that the critical nodes are at different levels. If two critical nodes are at the same level, then the value of one of the nodes is increased or decreased by a small value $\epsilon$. After applying these two steps for all Reeb graphs in the MDRG, we proceed to compute the MDPD.

\subsection{Shape Matching Dataset}
To compute a distance between two shapes, we first consider one or more scalar fields on each shape and construct a topological descriptor. To compare two shapes, we compute a distance between their topological descriptors. In this paper, we compute distances between shapes based on the scalar fields:  normalized geodesic distance ($\mu_n$) and normalized Euclidean distance ($d2_n$). We note, the geodesic distance between two points on a shape is defined as the length of the shortest path between them. Based on the geodesic distances between various points in a shape, $\mu_n$ is computed as described in \cite{hilaga2001topology}. The Euclidean distance between two points is the length of the line segment joining them. Similar to $\mu_n$, $d2_n$ for a shape is computed from Euclidean distances between various points of the shape. We note, to construct an MDRG, we require a JCN as input. We construct the JCN by subdividing the ranges of each of the component scalar fields into $32$ levels of quantization. In the experiments, we compute MDRGs of shapes for the bivariate field $(\mu_n, d2_n)$, where $\mu_n$ is taken as the first field. To analyze the effectiveness of the proposed method, we evaluate its performance as described below.

\textbf{Evaluation Measures: } The accuracy of a distance measure is measured based on the standard evaluation techniques Nearest Neighbour (NN), First Tier (FT), Second Tier (ST), E-measure, and Discounted Cumulative Gain (DCG). These techniques take a distance matrix consisting of the distances between all pairs of shapes as input and return a value between $0$ and $1$, where higher values indicate better accuracy. We provide a short description of these measures here and refer the readers to Shilane \etal \cite{2004-Shilane-Princeton-Shape-Benchmark}, for more details. NN is the percentage of the closest matches that belong to the same query class. FN and SN represent the percentage of the number of models belonging to the query class that appear within the top $K$ retrieved models. For FN, $K$ is set as $|C| - 1$ and for SN, $K$ is set as $2 \cdot(|C| - 1)$, where $|C|$ is the size of the query class. For each query model in class $C$ and any number of retrieved models $K$, precision represents the percentage of the number of retrieved models belonging to class $C$, while recall indicates the ratio of the number of models in class $C$ returned within the $K$ retrieved models. The E-measure (also known as $F_1$-score)  is a combination of precision and recall, defined as \cite{1975-Rijsbergen-Information-Retrieval,2003-Leifman-Signatures-3D-Models}:
\begin{footnotesize}
$$   E = \displaystyle\frac{2}{\frac{1}{\mathrm{Precision}} + \frac{1}{\mathrm{Recall}}}$$
\end{footnotesize}
where the number of retrieved models ($K$) is chosen as $32$ for computing the precision and recall. Finally, DCG gives more weightage to the retrieved models belonging to the query class which appear near the beginning of the list, compared to those that appear later in the list.

\subsubsection*{SHREC 2010 - Non-Rigid 3D Shape Retrieval}
\label{subsec:SHREC-2010}
\begin{figure}
    \centering
    \includegraphics[width=0.4\textwidth]{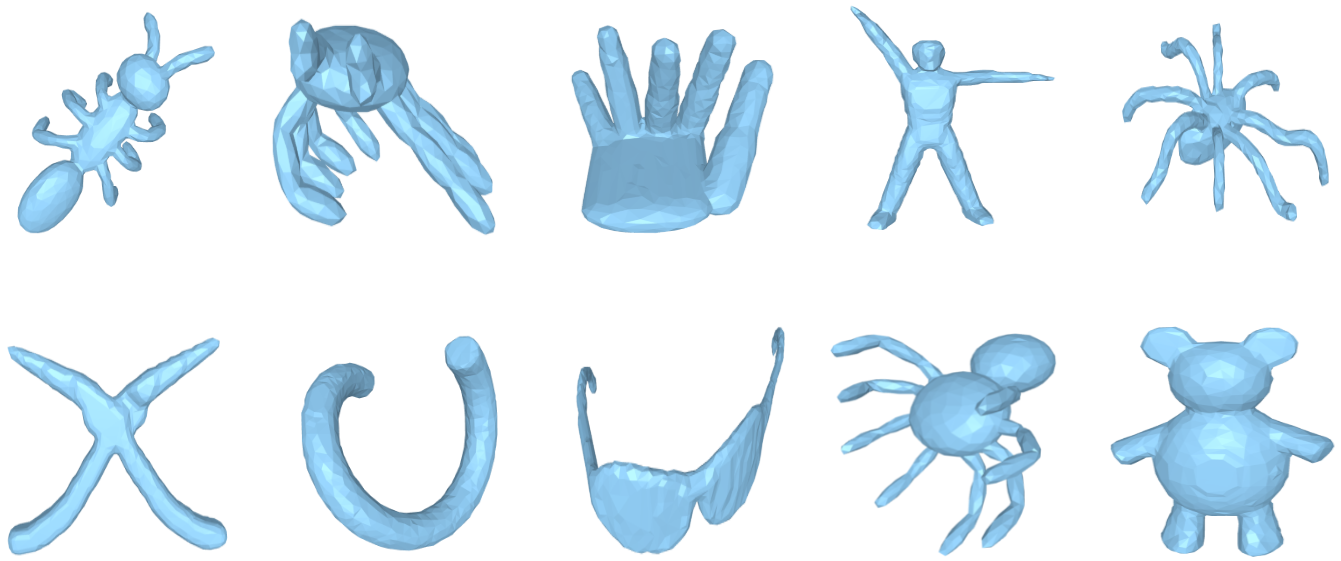}
    \caption{Collection of shapes from SHREC $2010$.}
    \label{fig:SHREC-2010-Shapes}
\end{figure}

The SHREC $2010$ dataset \cite{2010-SHREC} consists of $200$ watertight shapes classified into $10$ categories. A sample of the shapes is shown in \figref{fig:SHREC-2010-Shapes}. Each of the meshes is simplified into $2000$ faces \cite{2014-Li-Persistence-Based-Structural-Recognition}. First, we show the significance of the proposed distance between MDRGs of bivariate fields over the Wasserstein distance between Reeb graphs of scalar fields in differentiating between various classes of shapes in this dataset.

\begin{figure}
    \centering
    \includegraphics[width=0.48\textwidth]{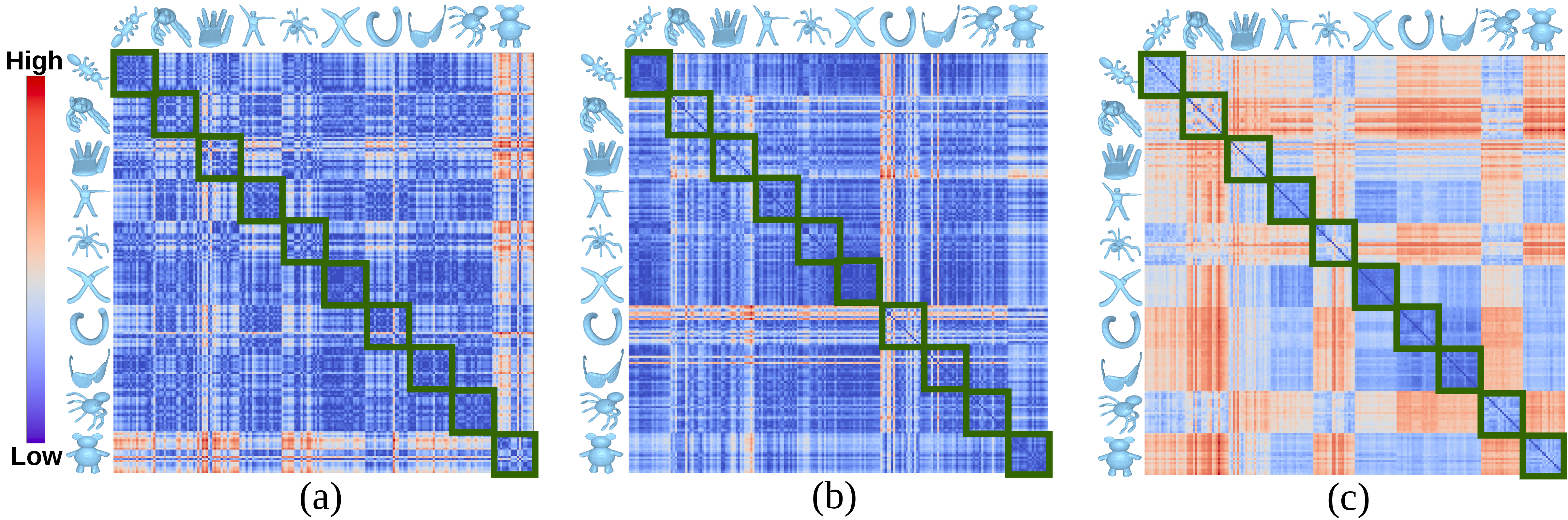}
    \caption{Wasserstein distance between all pairs of shapes in SHREC $2010$ dataset. (a) Distance matrix computed using Reeb graphs of $\mu_n$. (b) Distance matrix computed using Reeb graphs of $d2_n$. (c) Distance matrix computed using the MDRG of the bivariate field ($\mu_n, d2_n$).}
    \label{fig:distance_matrix_mdpd_shape_matching}
\end{figure}
\begin{table}
    \centering
    \caption{SHREC $2010$ dataset: Performance of the Wasserstein distance between Reeb graphs of scalar fields $\mu_n, d2_n$ and the Wasserstein distance between MDRGs of the bivariate field ($\mu_n, d2_n$).}
\label{table:wasserstien-distance-multi-field-significance}
\scalebox{0.9}{    \begin{tabular}{|c|c|c|c|c|c|}
        \hline
        \textbf{Fields} & \textbf{NN} & \textbf{$1$-Tier} & \textbf{$2$-Tier} & \textbf{e-Measure} & \textbf{DCG}\\
        \hline
        $\mu_n$ & $0.2100$ &  $0.2308$ & $0.3942$ & $0.2580$ & $0.5520$\\
        \hline
        $d2_n$ & $0.2700$ & $0.2039$ & $0.3484$ & $0.2290$ & $0.5444$\\
        \hline
        $(\mu_n, d2_n)$ & $0.9000$ & $0.6076$ & $0.8084$ & $0.5661$ & $0.8677$\\
        \hline
\end{tabular}}
\end{table}

\subsubsection*{Significance of Multi-field}
We compare the performance of Reeb graphs of scalar fields with MDRGs of bivariate fields using the Wasserstein distance. Table \ref{table:wasserstien-distance-multi-field-significance} shows the performance results of the proposed Wasserstein distance between MDRGs of the bivariate field ($\mu_n, d2_n$) and the Wasserstein distance between the Reeb graphs of the individual scalar fields $\mu_n$ and $d2_n$. \figref{fig:distance_matrix_mdpd_shape_matching} shows the distance matrices corresponding to the Wasserstein distance between all pairs of shapes based on scalar and bivariate fields. Each of the green boxes indicates distances between shapes belonging to a particular class. The effectiveness of using bivariate field over scalar fields in the retrieval of shapes can be clearly seen from the table and distance matrices.

\subsubsection*{Comparison with Other Methods}
We compare the proposed technique with the Wasserstein distance between persistence diagrams of MSMs \cite{2016-MultiscaleMapper}, approximate matching distance between bi-filtered JCNs \cite{2019-Kerber-Approximation-Matching-Distance}, distance between fiber-component distributions \cite{2019-Agarwal-histogram}, and distance between multi-resolution Reeb spaces \cite{2021-Ramamurthi-MRS}, for the bivariate field $(\mu_n,d2_n)$. For computing the matching distance, the maximal subdivision depth parameter is set to $4$. To construct a multiscale Mapper, the sequence of JCNs in the MRS is considered as a tower of simplicial complexes  (see \cite{2021-Ramamurthi-MRS} for more details). 
\begin{figure}
    \centering
    \includegraphics[width=0.48\textwidth]{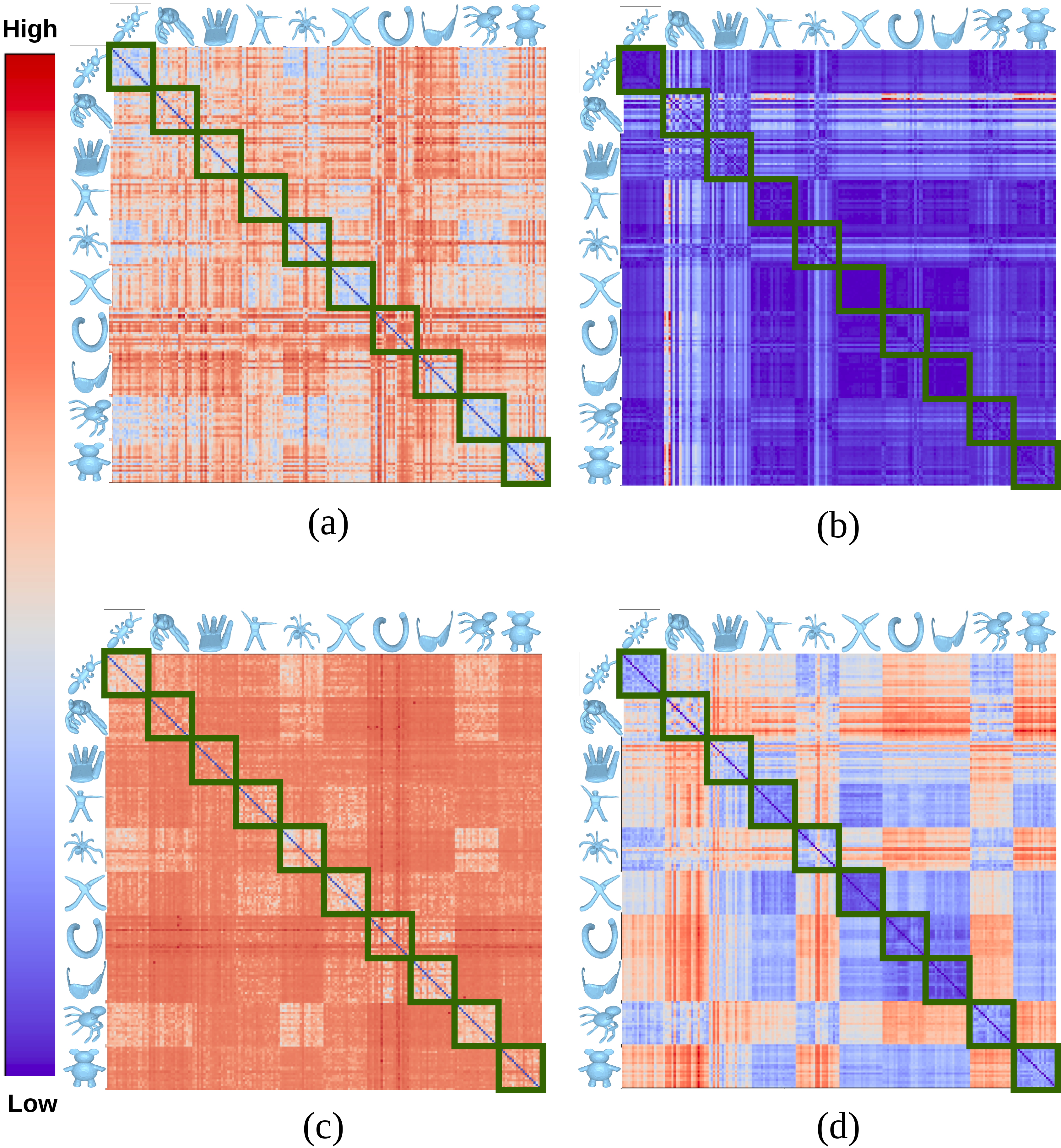}
    \caption{Comparison of Distance matrices for SHREC $2010$ dataset. (a) Distance between fiber-component distributions. (b) Wasserstein distance between persistence diagrams of MSMs. (c) Distance between MRSs. (d) Wasserstein distance between MDRGs.}
    \label{fig:distance-matrices-shrec-2010}
\end{figure}
The persistence diagram of the resulting tower is computed using the algorithm developed by Dey \etal \cite{2016-MultiscaleMapper, TamalDey2014TopologicalPersistenceForSimplicialMaps}. We compute the persistence diagrams and Wasserstein distances using the Sophia \cite{Reference-Sophia} and GUDHI \cite{gudhi:urm} libraries, respectively. Table \ref{table:comparison-shrec-2010} shows the results of retrieval accuracy. \figref{fig:distance-matrices-shrec-2010} shows the distance matrices computed using four of the methods. It can be observed from the table that the proposed technique has  better retrieval accuracy than the other methods. From the figure, we see that the proposed method outperforms other methods in discriminating between different classes of shapes. For instance, the method in \figref{fig:distance-matrices-shrec-2010}(a) involves projecting the Reeb space onto the range to obtain a fiber-component distribution, leading to the loss of topology. For the MSM in \figref{fig:distance-matrices-shrec-2010}(b), shapes in the $2^{nd}$ and $3^{rd}$ categories have relatively higher distances to other categories of shapes, and consequently, the remaining pairs of shapes are relatively closer as indicated by the color map. Similarly, for the MRS in \figref{fig:distance-matrices-shrec-2010}(c), the distance between any two identical shapes is very low compared to the distance between non-identical shapes, as indicated by dark blue color along the diagonal and red color in the rest of the matrix. In contrast, \figref{fig:distance-matrices-shrec-2010}(d) demonstrates such phenomena do not occur with the MDPD. 

Furthermore, in Figures \ref{fig:distance_matrix_mdpd_shape_matching}(c) or \ref{fig:distance-matrices-shrec-2010}(d), we observe that the  cross-shapes and U-shapes are relatively closer, as seen by the blue boxes of the $6^{th}$ row and $7^{th}$ column, and $7^{th}$ row and $6^{th}$ column. This is due to the fact that the topology of a U-shape is similar to that of a cross shape compared to other shapes.  However, the distance between two U-shapes is lower than the distance between a U-shape and a cross-shape, as indicated by the corresponding darker blue boxes along the diagonal of Figures 
 \ref{fig:distance_matrix_mdpd_shape_matching}(c) or \ref{fig:distance-matrices-shrec-2010}(d). Moreover, the proposed method captures not only the topological features of an MDRG but also their persistences.
\begin{table}
    \centering
    \caption{SHREC $2010$ dataset: Comparison of Wasserstein distance between persistence diagrams of MSMs \cite{2016-MultiscaleMapper}, approximate matching distance between bi-filtered JCNs \cite{2019-Kerber-Approximation-Matching-Distance}, distance between fiber-component distributions \cite{2019-Agarwal-histogram}, distance between MRSs \cite{2021-Ramamurthi-MRS}, and Wasserstein distance between MDRGs using the bivariate field $(\mu_n,d2_n)$.}
\label{table:comparison-shrec-2010}\scalebox{0.9}{
    \begin{tabular}{|c|c|c|c|c|c|}
        \hline
        \textbf{Methods} & \textbf{NN} & \textbf{$1$-Tier} & \textbf{$2$-Tier} & \textbf{e-Measure} & \textbf{DCG}\\
        \hline
        MSM & $0.5500$ & $0.3374$ & $0.5166$ & $0.3529$ & $0.6580$\\  
        \hline         
        Matching Distance & $0.1700$ & $0.1121$ & $0.2187$ &  $0.1288$  & $0.4715$\\
        \hline         
        Histogram & $0.6600$ & $0.3347$ & $0.4908$ &  $0.3325$  & $0.6874$\\
        \hline        
        MRS & $0.7600$  & $0.4634$ & $0.6682$  & $0.4596$ &   $0.7740$\\
        \hline
        MDRG & $0.9000$ & $0.6076$ & $0.8084$ & $0.5661$ & $0.8677$\\
        \hline
\end{tabular}}
\end{table}

\subsubsection*{Comparison with Spectral Descriptors}
We compare the performance of the proposed method  using the bivariate field $(\mu_n, d2_n)$ with the performances using the bivariate fields obtained from Heat Kernel Signature (HKS) \cite{2009-Sun-HKS} and Wave Kernel Signature (WKS) \cite{2011-Aubry-WKS}. The HKS at a point on the shape is defined as the amount of heat that remains after time $t$. In our experiment, we consider a bivariate field by computing the HKS corresponding to the first two values of $t$, in the discrete temporal domain. The WKS represents the average probabilities of measuring quantum particles of different energy levels at a point in the shape. In our experiment, we consider the WKS corresponding to the two smallest energy levels as a bivariate field. 

\begin{figure}[!h]
    \centering
    \includegraphics[width=0.48\textwidth]{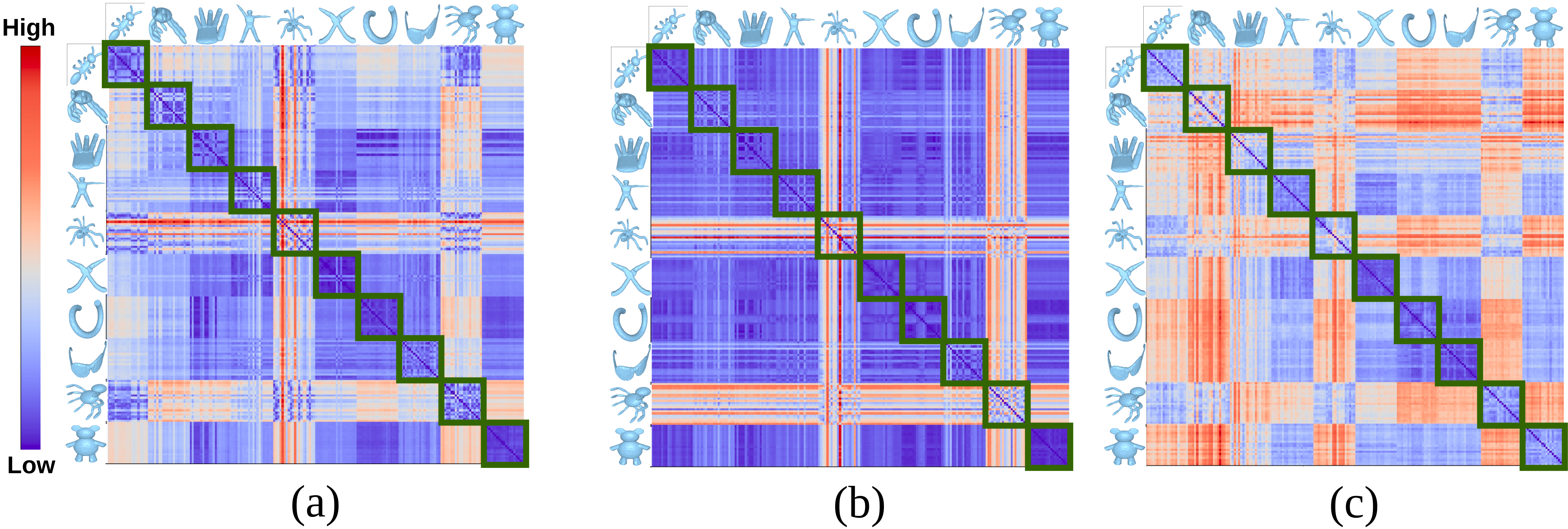}
    \caption{SHREC $2010$ dataset: Comparison of the bivariate fields obtained from HKS and WKS with the bivariate field $(\mu_n, d2_n)$, based on the Wasserstein distance between MDRGs. (a) Distance matrix computed using HKS. (b) Distance matrix computed using WKS. (c) Distance matrix computed using the bivariate field ($\mu_n, d2_n$).}
    \label{fig:distance_matrix_HKS_WKS}
\end{figure}

We assess the performance of HKS and WKS using the standard evaluation measures, and the results are presented in Table \ref{table:shape-matching-HKS-WKS}. Additionally, we visualize the corresponding distance matrices in \figref{fig:distance_matrix_HKS_WKS}. It can be seen that the proposed method using the bivariate field $(\mu_n,d2_n)$ is more effective than HKS and WKS.

\begin{table}
    \centering
    \caption{SHREC $2010$ dataset: Performance of the bivariate fields obtained from HKS, WKS, and the bivariate field $(\mu_n, d2_n)$, based on the Wasserstein distance between MDRGs.
   }
    \label{table:shape-matching-HKS-WKS}
\scalebox{0.9}{    \begin{tabular}{|c|c|c|c|c|c|}
        \hline
        \textbf{Fields} & \textbf{NN} & \textbf{$1$-Tier} & \textbf{$2$-Tier} & \textbf{e-Measure} & \textbf{DCG}\\
        \hline
        HKS & $0.8500$ & $0.5150$ & $0.7095$ & $0.4914$ & $0.8000$\\
        \hline
        WKS & $0.7700$ & $0.3871$ & $0.5421$ & $0.3745$ & $0.7336$\\
        \hline
        $(\mu_n, d2_n)$ & $0.9000$ & $0.6076$ & $0.8084$ & $0.5661$ & $0.8677$\\
        \hline
\end{tabular}}
\end{table}

In the next experiment, we demonstrate the performance of the proposed distance measure in a dataset from computational chemistry.

\subsection{Chemistry Data: Pt-CO Bond Formation}
\label{subsec:Pt-CO}
Adsorption refers to the phenomenon in which the molecules of a gas cling or bind onto the surface of a metal. This phenomenon has been applied in diverse fields such as corrosion, molecular electronics, electrochemistry, and heterogeneous catalysis \cite{2010-JACS-Kendric,book-Surface-Chem}. The adsorption of Carbon Monoxide (CO) on Platinum (Pt) surface has drawn attention, especially because of its significance in industrial sectors such as fuel cells, automobile emission, and other catalytic processes \cite{2009-dimakis-attraction, 2018-patra-surface}. Consequently, the investigation of the atomic-level interaction between the CO molecule and the platinum surface is of utmost importance.

\begin{figure}
    \centering
    \includegraphics[width=0.46\textwidth]{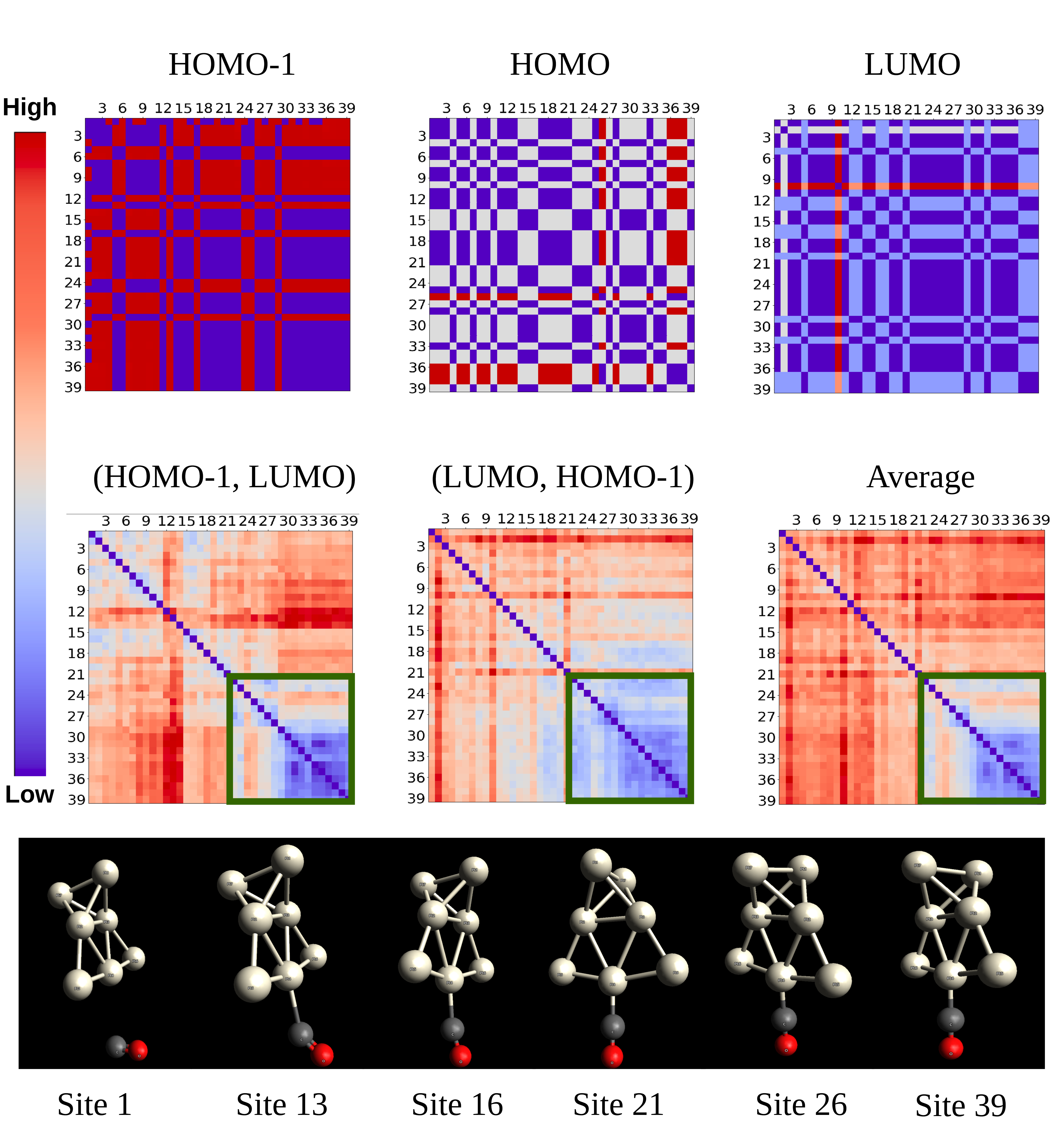}
    \caption{Distance matrices for Pt-CO data. The top row displays the distance computed using the scalar fields HOMO-$1$, HOMO, and LUMO. The first two matrices of the middle row are the distance matrices using the bivariate field  HOMO-$1$ and LUMO for two field orderings. The rightmost matrix is their average distance matrix. The bottom row shows the geometry of the Pt-CO bond formation at site $13$, whereas the stable bond formation happens at site $21$ \cite{2019-Agarwal-histogram}. This bond stabilization can be observed by a cluster of sites, highlighted by a green box in the matrices of the middle row. 
    }
    \label{fig:DFT-Scalar-Multi-Fields}
\end{figure}

The proposed method is validated in this research by examining the interaction between the CO molecule and a Pt$_7$ cluster. As the CO molecule nears the Pt$_7$ cluster, its internal bond weakens, resulting in the formation of a bond between the C atom of the CO molecule and one of the Pt atoms\cite{2009-dimakis-attraction}. The formation of the Pt-CO bond occurs at site $13$, which is confirmed by the geometry illustrated in \figref{fig:DFT-Scalar-Multi-Fields}. Nevertheless, the bond length at this site is unstable, and it stabilizes at site $21$ when the bond length between the Pt and C atoms reaches $1.84$\AA \cite{2019-Agarwal-histogram}. The objective of this experiment is to observe the formation of a stable bond between CO and Pt$_7$ molecules.

The Pt-CO dataset contains electron density distributions that were generated using quantum mechanical computations for the HOMO (Highest Occupied Molecular Orbital), LUMO (Lowest Unoccupied Molecular Orbital), and HOMO-$1$ orbitals. These distributions are defined on a regular grid of size $41 \times 41 \times 41$. The HOMO-$1$, HOMO, and LUMO orbitals are represented by orbital numbers $69, 70$, and $71$ respectively. The electron density distributions have been computed at $39$ different sites, with varying distances between the C atom of the CO molecule and the Pt-surface.

\subsubsection*{Observations and Results}
We evaluate the proposed measure's ability to detect the formation of a stable Pt-CO bond by analyzing different combinations of orbital densities at HOMO, LUMO, and HOMO-$1$ molecular orbitals. To construct the JCNs, we subdivide the range of each field into $4$ levels of quantization. For each field combination, we visualize a two-dimensional distance matrix by calculating the distance measure between every pair of sites. The top row of \figref{fig:DFT-Scalar-Multi-Fields} shows the distance matrices using the scalar fields HOMO, HOMO-$1$, and LUMO, respectively. However, none of these  matrices are able to show the location of the stable bond formation.

The second row of \figref{fig:DFT-Scalar-Multi-Fields} shows the distance matrices using the bivariate field HOMO-$1$ and LUMO, for different field orderings. The matrices for the field orderings (HOMO-$1$, LUMO) and (LUMO, HOMO-$1$) are represented in the left and the middle columns, respectively. Additionally, the rightmost matrix in the  second row shows the average of these distance matrices for two field orderings.
Notably, in all three matrices,  we observe a cluster containing sites $22$ to $39$  which corresponds to the stable bond formation.

\begin{figure}
    \centering
    \includegraphics[width=0.46\textwidth]{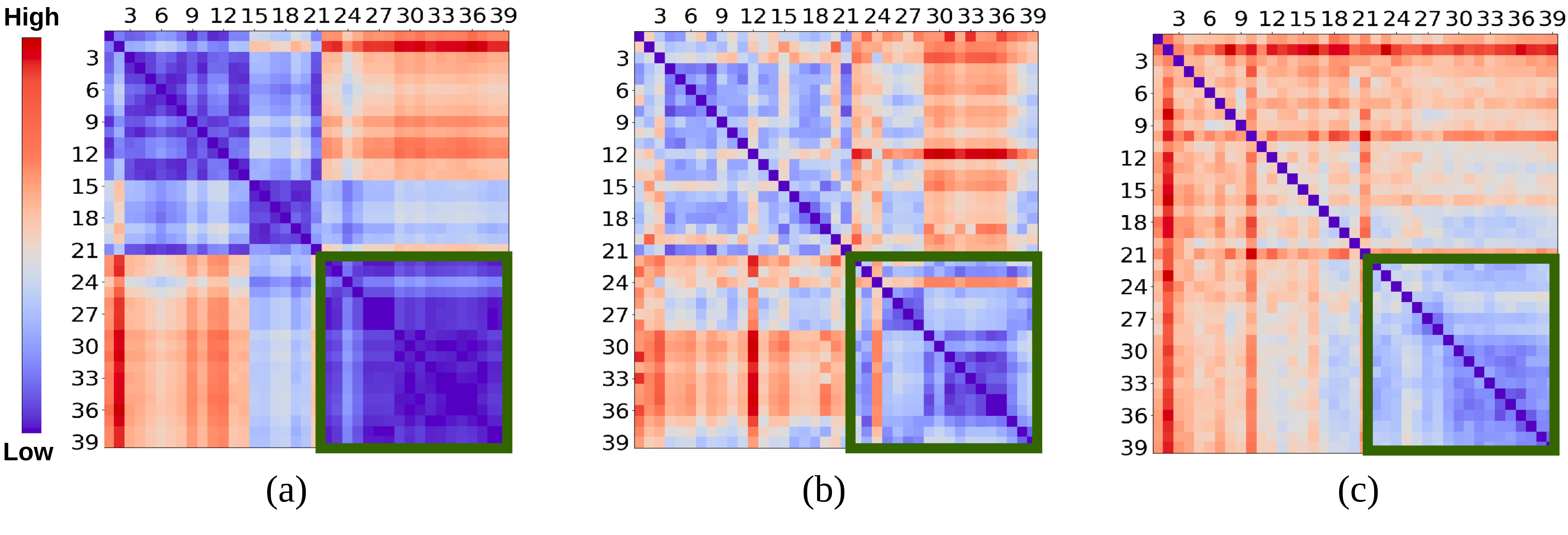}
    \caption{Distance matrices for the Pt-CO data obtained using (a) Wasserstein distance between  persistence diagrams of MSMs \cite{2016-MultiscaleMapper}, (b) Distance between MRSs \cite{2021-Ramamurthi-MRS}, and (c) proposed Wasserstein distance between MDRGs. For each matrix, a unique  cluster of sites that follow the bond stabilization is  highlighted by a green box.}
    \label{fig:DFT-Comparison-Multi-Field-Methods}
\end{figure}

We compare the performance of the proposed method with two other distance measures: (i) Wasserstein distance between persistence diagrams of MSMs \cite{2016-MultiscaleMapper} and (ii) distance based on MRSs  \cite{2021-Ramamurthi-MRS}. \figref{fig:DFT-Comparison-Multi-Field-Methods} shows the distance matrices based on MSMs, MRSs,  and the proposed MDPDs, respectively. 
We observe that a unique cluster of sites after stable bond formation is visible using the proposed distance measure as compared to the other distances that show more than one cluster.

\subsection{Computational Performance}
\label{subsec:computational-performance}
All the experiments were performed on a $2.20$ GHz $10$-Core Intel Xeon(R) with $16$ GB memory, running Ubuntu $16.04$. Table \ref{table:RuntimeStat} shows the computational performance of the proposed distance measure between a pair of shapes from SHREC $2010$ and a pair of sites from Pt-CO dataset, using scalar and bivariate fields. The table  shows the time taken for constructing two MDPDs from the corresponding JCNs and to compute the Wasserstein distance between them.
\begin{table}
    \caption{Computational performance results for SHREC $2010$ and Pt-CO datasets. Here, Shapes/Sites: Shapes/sites compared; Levels: quantization levels of the JCNs; Field(s): Component scalar fields for constructing the MDRGs; $\abs{V_1},\abs{V_2}$: Number of vertices in the JCNs; $T_{\mathrm{MDPD}}$: Time taken (in seconds) for constructing the MDPDs from the JCNs, and $T_{d_{W,q}}$: the time taken (in seconds) for computing the Wasserstein distance between the MDRGs.}
    \centering
    \resizebox{\columnwidth}{!}{%
    \begin{tabular}{|c|c|c|c|c|c|c|c|}
        \hline
         \textbf{Dataset} & \textbf{Shapes/Sites} &  \textbf{Levels} & \textbf{Field(s)} & $\abs{V_1},\abs{V_2}$ & $T_{\mathrm{MDPD}}$ & $T_{d_{W,q}}$\\ \hline
         \multirow{3}{*}{SHREC $2010$}  & \multirow{3}{*}{Human, Hand} & \multirow{3}{*}{$32, 32$} & $\mu_n$ & $92, 103$ & $0.0031$s & $0.0001$s\\ \cline{4-7}
         & & & $d2_n$ & $107, 111$ & $0.0030$s & $0.0001$s\\ \cline{4-7}
         & & & $\mu_n, d2_n$ & $380, 1037$ & $0.0635$s & $0.0059$s\\ \hline          
         \multirow{3}{*}{Pt-CO}  & \multirow{3}{*}{$21, 22$}  & \multirow{3}{*}{$4, 4$} & HOMO-$1$ & $37, 32$ & $0.0023$s & $0.0001$s\\ \cline{4-7}
         & & & LUMO & $55, 27$ & $0.0042$s & $0.0001$s\\ \cline{4-7}
         & & & HOMO-$1$, LUMO & $417, 245$ & $0.0402$s & $0.0031$s\\ \hline    
    \end{tabular}}
\label{table:RuntimeStat}
\end{table}

Table \ref{tab:RuntimeStat-comparison-of-methods} shows the comparison of the running times of the proposed distance measure with the Wasserstein distance between persistence diagrams of MSMs \cite{2016-MultiscaleMapper}, approximate matching distance between bi-filtered JCNs \cite{2019-Kerber-Approximation-Matching-Distance}, distance between fiber-component distributions \cite{2019-Agarwal-histogram}, and the distance between MRSs \cite{2021-Ramamurthi-MRS}. We note, except for the approximate matching distance, all other methods involve the computation of JCNs which costs most of the computational time. For the  SHREC $2010$ dataset, the computation of pair of JCNs takes approximately $0.8$ seconds, and for the Pt-CO dataset, it  takes approximately $78$ seconds. We observe, in general, the proposed technique is computationally efficient as compared to other methods.

\begin{table}
    \centering
    \caption{Comparison of the computational performance results for the proposed distance between MDPDs with various other methods for SHREC $2010$ and Pt-CO datasets. Here, Shapes/Sites: Shapes or sites compared; Levels: quantization levels of the JCNs; Field(s): Component scalar fields for constructing the MDRGs; Method: Method used for comparison of data, and Time: time (in seconds) for computing the distance between shapes/sites (including the time for computing JCNs).}
    \label{tab:RuntimeStat-comparison-of-methods}    
    \resizebox{\columnwidth}{!}{%
    \begin{tabular}{|c|c|c|c|c|c|c|}
       \hline   
       \textbf{Dataset}  & \textbf{Shapes/Sites} & \textbf{Levels} & \textbf{Fields} & \textbf{Method} & \textbf{Time} \\\hline
       \multirow{5}{*}{SHREC $2010$} & \multirow{5}{*}{Human, Hand} & \multirow{5}{*}{$32,32$} & \multirow{5}{*}{$\mu_n, d2_n$} & MSM & $2.8423$s\\\cline{5-6}
       & & & & Approx. Matching Distance & $56.0083$s\\\cline{5-6}
       & & & & Histogram & $0.8860$s\\\cline{5-6}
       & & & & MRS & $1.4238$s\\\cline{5-6}      
       & & & &  MDPD & $1.0205$s\\\hline
       \multirow{5}{*}{Pt-CO} & \multirow{5}{*}{$21, 22$} & \multirow{5}{*}{$4, 4$} & \multirow{5}{*}{HOMO-$1$, LUMO} & MSM & $79.1865$s\\\cline{5-6}
       & & & & Approx. Matching Distance & $130.5855$s\\\cline{5-6}
       & & & & Histogram & $78.8008$s\\\cline{5-6}
       & & & & MRS & $78.2939$s\\\cline{5-6}
       & & & & MDPD & $78.5958$s\\\hline
    \end{tabular}}
\end{table}

\subsection{Effect of JCN Quantization Levels}
\label{subsec:varying-quantization-levels}
\begin{figure}
    \centering
    \includegraphics[width=0.4\textwidth]{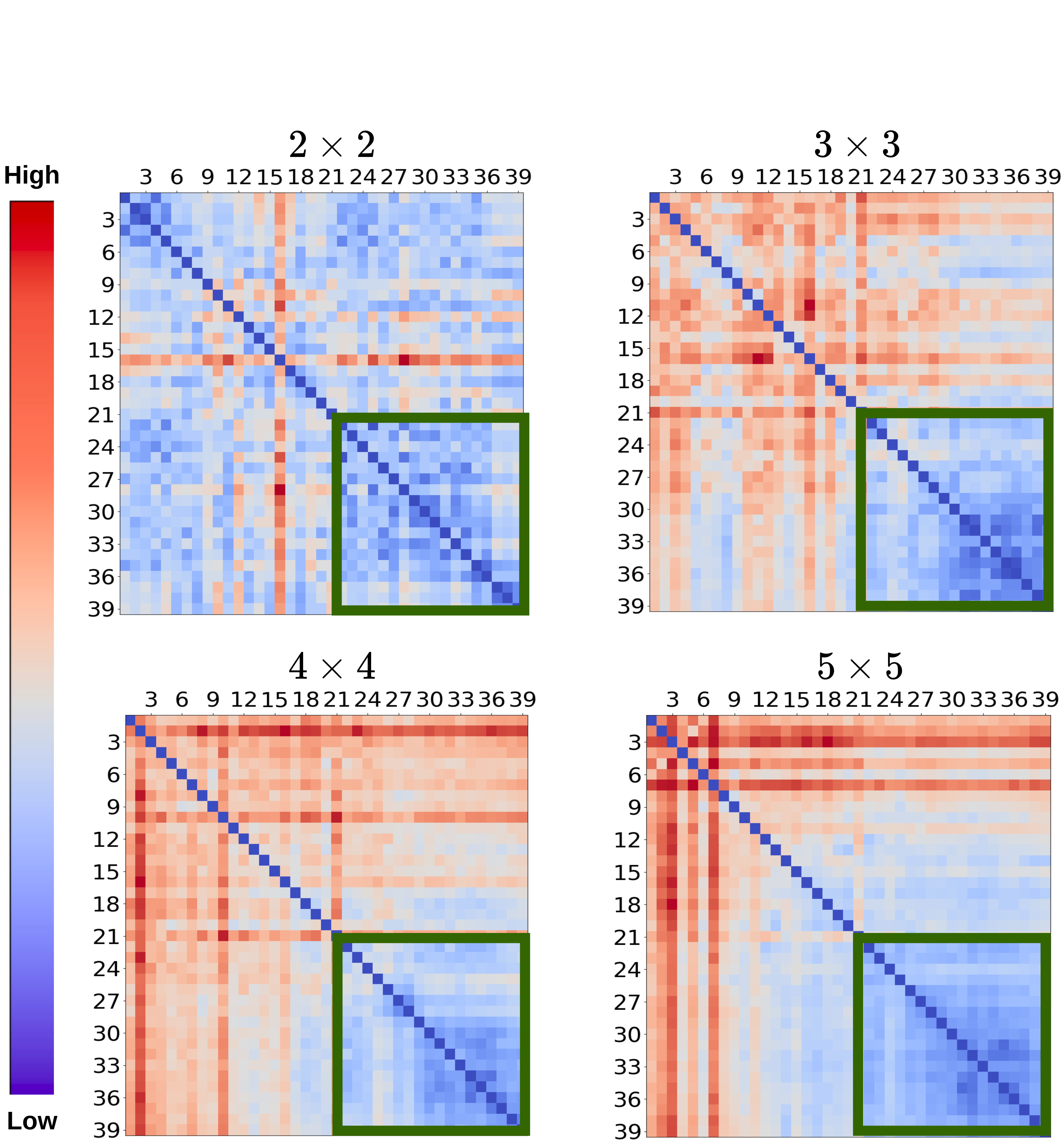}
    \caption{Distance matrices computed using the Wasserstein distance between MDRGs for all pairs of sites in the Pt-CO data using the bivariate field (LUMO, HOMO-$1$)  for $2 \times 2$, $3 \times 3$, $4 \times 4$, and $5 \times 5$ quantization levels. The cluster (sites $22$-$39$) after bond stabilization at site $21$ is indicated by a green box, in each of the matrices.} 
    \label{fig:DFT-varying-slabs}
\end{figure}

In this subsection, we discuss the effect of JCN quantization levels on the performance of the proposed method. Table \ref{table:performance-shape-matching-varying-slabs} presents the performance results of the proposed method on the SHREC $2010$ dataset for $5$ different quantization levels of the bivariate field $(\mu_n, d2_n)$ (as discussed in \secref{subsec:SHREC-2010}). \figref{fig:DFT-varying-slabs} shows the distance matrices for the Pt-CO data corresponding to four different quantization levels of the bivariate field consisting of LUMO and HOMO-$1$ (as in \secref{subsec:Pt-CO}). We observe an improvement in the performance of the proposed method with an increase in quantization levels, in both datasets. However, for choosing the levels of quantization one needs to find a trade-off between the performance results and the computational time. Moreover, when the number of quantization levels is sufficient to capture the topology of the Reeb space, then further increasing the quantization levels would not have a significant impact on the distance. For instance, the performance results for the SHREC $2010$ dataset using  $16\times 16$ and $32 \times 32$ quantization levels are nearly the same (see Table \ref{table:performance-shape-matching-varying-slabs}). Similarly, for the Pt-CO dataset, the cluster of sites after bond stabilization is visible using quantization levels $3 \times 3$, $4 \times 4$, and $5 \times 5$ (see \figref{fig:DFT-varying-slabs}).

\begin{table}
    \centering
    \caption{SHREC $2010$ dataset: Performance of the Wasserstein distance between MDRGs for all pairs of shapes using the bivariate field $(\mu_n, d2_n)$ for $5$ different quantization levels.}
    \label{table:performance-shape-matching-varying-slabs}
\scalebox{0.9}{    \begin{tabular}{|c|c|c|c|c|c|}
        \hline
        \textbf{Levels} & \textbf{NN} & \textbf{$1$-Tier} & \textbf{$2$-Tier} & \textbf{e-Measure} & \textbf{DCG}\\
        \hline
        $2 \times 2$ & $0.5650$ & $0.3971$ & $0.6342$ & $0.4247$ & $0.7087$\\
        \hline
        $4 \times 4$ & $0.7650$ & $0.4655$ & $0.6903$ & $0.4598$ & $0.7702$\\
        \hline
        $8 \times 8$ & $0.8400$ & $0.5321$ & $0.7587$ & $0.5253$ & $0.8270$\\
        \hline
        $16 \times 16$ & $0.8950$ & $0.5824$ & $0.7866$ & $0.5488$ & $0.8542$\\    
        \hline
        $32 \times 32$ & $0.9000$ & $0.6076$ & $0.8084$ & $0.5661$ & $0.8677$\\
        \hline
\end{tabular}}
\end{table}

\section{Conclusions and Future Works}
\label{sec:conclusions}
In this paper, we have proposed the multi-dimensional persistence diagram corresponding to an MDRG of a quantized Reeb space. 
Based on this, we have developed a novel distance measure between two multi-fields by extending the Wasserstein distance between two Reeb graphs. 
The proposed distance measure is a pseudo-metric and is shown to satisfy a stability property.
We observed the discriminating power of the proposed method in the SHREC $2010$ dataset.
Its effectiveness is also shown in a computational chemistry dataset through the detection of the stable bond formation between Pt and CO molecules.
The proposed method is seen to perform   better as compared to other related topological techniques in the literature.

However, computing the multi-dimensional persistent features  of a Reeb space  dispensing with the quantization of the joint contour net \cite{2016-Julien-ReebSpace}  is a challenging open problem.
We note, the MDRG and consequently the MDPD structures corresponding to a multi-field depend on the field orderings, although the overall topology of the Reeb space, they capture, remains the same. However, to define a topological distance between two multi-fields or Reeb spaces that is independent of the field orderings one needs to find a generalization of the persistence diagram for a multifiltration which is an open problem. Furthermore, the worst-case time complexity of computing the Wasserstein distance using the Hungarian algorithm is cubic on the number of points in the persistence diagrams. One can try using the Auction algorithm \cite{2017-Kerber-Auction-Algorithm} to improve the performance which should be addressed in a future work.

\ifCLASSOPTIONcompsoc
  \section*{Acknowledgments}
\else
  \section*{Acknowledgment}
\fi

The authors would like to thank the Science and Engineering Research Board (SERB), India (SERB/CRG/2018/000702) and International Institute of Information Technology (IIITB), Bangalore for funding this project and for generous travel support.

\ifCLASSOPTIONcaptionsoff
  \newpage
\fi




%

\bibliographystyle{IEEEtran}
\bibliography{template.bib}




\appendices
\section{Notations}
\label{appendix:notations}
\newcolumntype{C}[1]{>{\centering\arraybackslash}m{#1}}

\begin{table}[H]
    \centering
    \renewcommand{\arraystretch}{2.2}    
    \caption{Important notations}
    \label{table:notations}
\scalebox{0.9}{    \begin{tabular}{|C{2cm}|m{6.5cm}|}
        \hline
        \textbf{Notation} & \textbf{Description}\\
        \hline
        $\RG_{f}$ & Reeb Graph of a scalar field $f$\\        
        \hline
        $\Dg_0(\RG_f)$ & $0$-th ordinary persistence diagram of $\RG_{f}$\\
        \hline
        $\ExDg_0(\RG_{f})$ & $0$-th extended persistence diagram of $\RG_{f}$\\
        \hline
        $\PD_0(\RG_f)$ & Union of $\Dg_0(\RG_f)$ and $\ExDg_0(\RG_f)$, excluding the point with infinite persistence in  $\Dg_0(\RG_f)$\\
        \hline
        $\ExDg_1(\RG_{f})$ & $1$-st extended persistence diagram of $\RG_{f}$\\
        \hline        
        $\PD(\RG_{f})$ & Persistence diagram of $\RG_{f}$ (obtained by taking the union of $\PD_0(\RG_f)$  and  $\ExDg_1(\RG_{f}))$\\
        \hline
        $\RS_{\f}$ & Reeb Space of a multi-field $\f$\\
        \hline
        $\JCN_{\f}$ & Joint Contour Net of a multi-field $\f$\\        
        \hline
        $\MDRG_{\f}$ & Multi-Dimensional Reeb Graph of a multi-field $\f$\\
        \hline
        $\RG_{\widetilde{f_2^p}}$ & Reeb graph of the restricted field $\widetilde{f_2^p}\equiv f_2|_{C_p}: C_p\rightarrow \mathbb{R}$, where $C_p:=q_{f_1}^{-1}(p)$ is the contour of $f_1$ (the first field) corresponding to the node $p \in \RG_{f_1}$\\
        \hline
        $(\gamma_1, \gamma_2^{p})$ & Pair of persistent homology classes corresponding to filtrations of $\RG_{f_1}$ and $\RG_{\tilde{f_2^{p}}}$, respectively\\
        \hline        
        $\PD^{p}_{\ux_1}(\MDRG_\f)$ & MDPD of $\MDRG_\f$ relative to a birth-death point $\ux_1\in \PD(\RG_{f_1})$ and a node $p\in \RG_{f_1}$\\
        \hline
        $\pI(\ux_1)$ &  Persistence interval of a birth-death point $\ux_1$ in a persistence diagram\\
        \hline
        $\PD^{\pI(\ux_1)}_{\ux_1}(\MDRG_\f)$ & MDPD of $\MDRG_\f$ relative to a birth-death point $\ux_1\in \PD(\RG_{f_1})$ and persistence interval $\pI({\ux_1})$\\
        \hline        
        $\PD(\MDRG_{\f})$ & Multi-Dimensional Persistence Diagram of $\MDRG_{\f}$\\ 
        \hline
        $\PD^{p}(\MDRG_\f)$ & Subset of $\PD(\MDRG_\f)$ relative to a node $p\in \RG_{f_1}$\\        
        \hline
        $d_{W,q}(\MDRG_{\f}, \MDRG_{\g})$ & Wasserstein distance between $\MDRG_{\f}$ and $\MDRG_{\g}$\\  
        \hline        
\end{tabular}}
\end{table}

Table \ref{table:notations} presents a list of key notations utilized throughout this paper.
\section{Persistence Diagrams}
\label{appendix:persistence-diagrams}
In this section, we give an introduction to persistent homology and refer the reader to \cite{book-herbert-computational-topology,2005-Zomorodian-ComputingPersistentHomology} for further details. Let $f: \cM \rightarrow \R$ be a continuous function. The sublevel set corresponding to a real number $a$ is defined as the set of points having the value of $f$ at most $a$, i.e. $\cM_{\leq a} = f^{-1}(-\infty, a]$. For every $a \leq b$, there is an inclusion map from $\cM_{\leq a}$ to $\cM_{\leq b}$ which induces a map $f_k^{a,b} : H_k(\cM_{\leq a}) \rightarrow H_k(\cM_{\leq b})$ between the corresponding $k$-homology groups.  A real number $a$ is said to be a \emph{homological critical value} of $f$ if there exists an integer $k \in \Z^{*}$ (where $\Z^{*}$: the set of non-negative integers) such that for every small $\delta > 0$, the map $f_{k}^{a-\delta,a+\delta}$ is not an isomorphism. We assume that $f$ is a \emph{tame} function, i.e $f$ has finite number of homological critical values $ \min f=a_1< \ldots < a_N=\max f$ and $\forall k \in \Z^{*}$, the homological groups $H_k(\cM_{\leq a})$ are of finite dimension. In this paper, we consider the component scalar fields of a multi-field to be tame functions and the homology with coefficients in $\Z_2$. Thus, for every $a_i$ with $0 \leq i \leq N$, $H_k(\cM_{\leq a_i})$ is a vector space and for $-\infty=a_0<a_1< \ldots < a_N $ we obtain a sequence of vector spaces:
\begin{footnotesize}
\begin{equation}
    \emptyset = H_k(\cM_{\leq a_0}) \rightarrow H_k(\cM_{\leq a_1}) \rightarrow \cdots \rightarrow H_k(\cM_{\leq a_N}) =  H_k(\cM).
    \label{eqn:homology-sequence-ordinary}
\end{equation}
\end{footnotesize}

\noindent
A homology class $\gamma$ is said to be born at $a$ if $\gamma \in H_k(\cM_{\leq a})$ but $\forall \delta > 0$ $\gamma \notin H_k(\cM_{\leq a-\delta})$. Similarly, a homology class $\gamma$ born at $a$ is said to die at $b$ if for any $\delta >0$, $f_k^{a,b-\delta}(\gamma) \notin \mathrm{Im} f_k^{a-\delta, b - \delta}$ but  $f_k^{a,b}(\gamma) \in \mathrm{Im}f_k^{a-\delta,b}$. Persistent homology records such events of birth and death of homology classes. The $k$-th ordinary persistence diagram encodes the births and deaths of $k$-dimensional homology classes as a multiset of points in $\overline{\R}^2$. Here, $\overline{\R} = \R \cup \{-\infty, \infty\}$. A point $(a,b)$ in the $k$-th ordinary persistence diagram corresponds to a $k$-dimensional homology class with birth $a$ and death $b$. The multiplicity of a point $(a,b)$ with $a < b$ is the number of homology classes born at $a$ and die at $b$. Points which lie on the diagonal, i.e. points of the form $(a,a)$, have infinite multiplicity. We will now see the homology classes which are born but do not die in the sequence in \eqnref{eqn:homology-sequence-ordinary}.

A homology class is said to be non-trivial or essential if it persists throughout the sequence of homology groups corresponding to a filtration. A $k$-dimensional essential homology class born at $a_i$ is captured by the point $(a_i,\infty)$ in the $k$-th ordinary persistence diagram. However, extended persistence captures both the birth and death of essential homology classes. We have a sequence of homology groups going up and a sequence of relative homology groups coming down as follows.
\begin{footnotesize}
\begin{equation}
\begin{aligned}
&\emptyset = H_k\left(\cM_{\leq a_0}\right) \rightarrow  H_k\left(\cM_{\leq a_1}\right) \rightarrow \cdots \rightarrow  H_k\left(\cM_{\leq a_N}\right)=  H_k\left(\cM\right)\\
&=H_k\left(\cM, \cM_{\geq a_N}\right) \rightarrow H_k\left(\cM, \cM_{\geq a_{N-1}}\right) \rightarrow \cdots \rightarrow H_k\left(\cM, \cM_{\geq a_0}\right) = \emptyset.
\end{aligned}
\label{eqn:homology-sequence-essential}
\end{equation}
\end{footnotesize}

\noindent where, $\cM_{\geq a_i}$ is defined as the set of points in $\cM$ with $f$ value at least $a_i$, $\cM_{\geq a_i} = f^{-1}[a_i,\infty)$. An essential homology class of dimension $k$ is born during the ordinary part of the sequence and dies in the relative part of the sequence in \eqnref{eqn:homology-sequence-essential}. Each essential class born at $H_k\left(\cM_{\leq a_i}\right)$ and dies at $ H_k\left(\cM, \cM_{\geq a_j}\right)$ is encoded by a point $(a_i,a_j)$ in the $k$-th extended persistence diagram. We note, the death time of an essential homology class is lesser than or equal to the birth time.

In the current paper, we construct the multi-dimensional persistence diagram of an MDRG from the persistence diagrams of its component Reeb graphs.
\section{Persistence Diagram of a Reeb Graph}
\label{appendix:reeb-graph-persistence-diagram}
In particular, the quotient space corresponding to a scalar field $f:\cM\rightarrow \R$ is a graph consisting of nodes and edges and is called the Reeb graph, denoted by $\RG_f$. Being a graph structure, the persistent homology features of a Reeb graph can be encoded in $0$- and $1$-dimensional persistence diagrams~\cite{2014-bauer-ReebGraph}. Let $\Dg_0(\RG_f)$ be the $0$-dimensional persistence diagram. Similarly, let $\ExDg_0(\RG_f)$ and $\ExDg_1(\RG_f)$ be, respectively, the $0$- and $1$-dimensional extended persistence diagrams. $\Dg_0(\RG_f)$ captures the $0$-dimensional persistent homology features, i.e. birth-death pairs occurring in the sublevel set filtration. Similarly, for capturing the birth-death of the loop-features we consider  $\ExDg_1(\RG_f)$. Furthermore, for Reeb graphs $\ExDg_0(\RG_f)$ captures the range of the function $\bar{f}$. For simplicity, we consider $f$ as a Morse function, i.e. all its critical points are non-degenerate and are at different levels. Then all critical nodes of $\RG_f$ have distinct $\bar{f}$ values and can be either of the following four-types: (i) a minimum (with down-degree~$=0$), (ii) a maximum (with up-degree~$=0$), (iii) a  down-fork (with down-degree~$=2$), and (iv) a up-fork (with up-degree~$=2$). Other nodes of degree $2$ with both up-degree and down-degree equal to $1$ are regular nodes. 
 
Consider the homology classes of the sublevel sets $(\RG_f)_{\leq a}=\bar{f}^{-1}(-\infty, a]$ for an increasing sequence of values of $a$. A $0$-dimensional homology class of $(\RG_f)_{\leq a}$ is born when $a$ passes through a critical value corresponding to a local minimum of $\bar{f}$. Now a down-fork (similarly, up-fork) node is called an \emph{essential down-fork} node when it contributes to a loop of the Reeb graph. Otherwise it is called an \emph{ordinary down-fork} node. Both up-forks nodes and down-forks nodes correspond to saddles of $\bar{f}$. At an ordinary down-fork  node $s$ two $0$-dimensional homology classes of $(\RG_f)_{< s}$ join to form a single class, i.e one of the previously born classes dies at node $s$.  Thus the class at ordinary down-fork node $s$ can be paired with a class which is born most recently at a minimum node $t$ and this corresponds to a point $(\bar{f}(t), \bar{f}(s))$ in the ordinary persistence diagram $\Dg_0(\RG_f)$ (see \figref{fig:reeb-persistence}(b)). Similarly, for $-f$, we can pair each ordinary up-fork node $u$ with a maximum node $v$  which is represented by a point $(\bar{f}(u), \bar{f}(v))$ in the persistence diagram $\Dg_0(\RG_{-f})$. The global minimum is paired with the global maximum and is represented as a point in the $0$-th extended persistence diagram $\ExDg_0(\RG_f)$ (see \figref{fig:reeb-persistence}(c)).

We use the notation $\PD_0(\RG_f)$ to denote the union of $\Dg_0(\RG_f)$ and $\ExDg_0(\RG_f)$, excluding the point with infinite persistence in  $\Dg_0(\RG_f)$. For \figref{fig:reeb-persistence}(d), $\PD_0(\RG_f):=\Dg_0(\RG_f)\cup \ExDg_0(\RG_f)\setminus \{(1,\infty)\}$. We are interested in capturing persistent features of $\RG_{f}$ whose birth and death are within the range of $\bar{f}$. The point $(1,\infty)$ captures only the global minimum of $\bar{f}$. However, the point $(1,12)$ in $\ExDg_0(\RG_f)$ captures the entire range of $\bar{f}$, i.e. it captures both the global minimum and maximum of $\bar{f}$. Therefore, we define $\PD_0$ by considering the points in $\Dg_0(\RG_{f})$ other than $(1,\infty)$, instead we consider the point $(1,12)$ in $\ExDg_0(\RG_{f})$.

To capture loops in the Reeb graph, each essential down-fork $p$ is matched with the corresponding essential up-fork $q$ which gives a point $(\bar{f}(p),\bar{f}(q))$ in the $1$-st extended persistence diagram $\ExDg_1(\RG_f)$. \figref{fig:reeb-persistence}(e) shows the $1$-st extended persistence diagram of the Reeb graph in \figref{fig:reeb-persistence}(a). Finally, to capture all the features of a Reeb graph in a single persistence diagram, we take the union of $\PD_0(\RG_{f})$ and $\ExDg_1(\RG_f)$, and denote it by $\PD(\RG_{f})$ (see \figref{fig:reeb-persistence}(f)). We use this representation for computing the multi-dimensional persistence diagram  in \secref{sec:mdpd}.
\section{Pseudo-Metric: Proof of Theorem 4.1}
\label{appendix:pseudo-metric}
\textbf{Identity: } Note that different Reeb graphs can have the same persistence diagram \cite{2014-bauer-ReebGraph}. Similarly, the MDPDs of two different MDRGs $\MDRG_{\f}$ and $\MDRG_{\g}$ can be the same. Therefore, $d_{W,q}(\MDRG_{\f}, \MDRG_{\g}) = 0$ does not imply that $\MDRG_{\f}$ and $\MDRG_{\g}$ are identical. Therefore, the identity property is not satisfied.

\noindent
\textbf{Non-negativity: } For any two MDRGs $\MDRG_{\f}$ and $\MDRG_{\g}$, we have $d_{W,q}(\MDRG_{\f}, \MDRG_{\g}) \geq 0$. Thus the non-negative property of $d_{W,q}$ holds.

\noindent
\textbf{Symmetry: } For $\MDRG_{\f}$ and $\MDRG_{\g}$, it is easy to see that $d_{W,q}(\MDRG_{\f}, \MDRG_{\g})$ and $d_{W,q}(\MDRG_{\g}, \MDRG_{\f})$ are equal. Thus the symmetry property is satisfied.

\noindent
\textbf{Triangle Inequality: } To show the triangle inequality, we need to prove that for any three MDRGs $\MDRG_{\f}, \MDRG_{\g}$ and $\MDRG_{\h}$, the following inequality holds:
\begin{footnotesize}
\begin{align*}
d_{W,q}(\MDRG_{\f},\MDRG_{\h})
\leq d_{W,q}(\MDRG_{\f},\MDRG_{\g}) +\\ d_{W,q}(\MDRG_{\g},\MDRG_{\h}).
\end{align*}
\end{footnotesize}

\noindent We first prove the triangle inequality for $q=1$. Let $\eta' : \PD(\MDRG_{\f}) \rightarrow \PD(\MDRG_{\g})$ be the optimal bijection between $\PD(\MDRG_{\f})$ and $\PD(\MDRG_{\g})$ in $d_{W,q}(\MDRG_{\f},\MDRG_{\g})$. Similarly, let $\eta'' : \PD(\MDRG_{\g}) \rightarrow \PD(\MDRG_{\h})$ be the optimal bjiection in $d_{W,q}(\MDRG_{\g}, \MDRG_{\h})$. Let $\eta''' = \eta'' \circ \eta'$ . Since $\eta'$ and $\eta''$ are bijections corresponding to the Wasserstein distance between MDRGs and $\eta''' = \eta'' \circ \eta'$, it follows that $\eta'''$ is a bijection between $\PD(\MDRG_{\f})$ and $\PD(\MDRG_{\h})$ satisfying the criteria (C1)-(C3) as mentioned in \secref{subsec:distance-between-MDRGs}. We now obtain a bound on $d_{W,q}(\MDRG_{\f}, \MDRG_{\h})$ as follows.
\begin{footnotesize}
\begin{align*}
&d_{W,q}(\MDRG_{\f},\MDRG_{\h})\\
&= \displaystyle\inf_{\eta: \PD(\MDRG_\f)\rightarrow
    \PD(\MDRG_\h)}\sum_{\x\in \PD(\MDRG_\f)}\|\x-\eta(\x)\|_{\infty}\\
    &\leq \sum_{\x\in \PD(\MDRG_\f)}\|\x-\eta'''(\x)\|_{\infty}\\
    &\leq \sum_{\x\in \PD(\MDRG_\f)}\left(\|\x-\eta'(\x)\|_{\infty} + \|\eta'(\x)-\eta''(\eta'(\x))\|_{\infty}\right)\\
    &= \sum_{\x\in \PD(\MDRG_\f)} \|\x-\eta'(\x)\|_{\infty} +\\
    &\indent \sum_{\x\in \PD(\MDRG_\f)}  \|\eta'(\x)-\eta''(\eta'(\x))\|_{\infty}\\
    &= \sum_{\x\in \PD(\MDRG_\f)} \|\x-\eta'(\x)\|_{\infty} +
    \sum_{\y\in \PD(\MDRG_\g)} \|\y-\eta''(\y)\|_{\infty}\\
    &\hspace{4cm}\text{ (since $\eta'$ is a onto map)}\\
    &=d_{W,q}(\MDRG_{\f}, \MDRG_{\g}) +
     d_{W,q}(\MDRG_{\g}, \MDRG_{\h}).
\end{align*}
\end{footnotesize}

\noindent For $q>1$, the triangle inequality can be proved similarly using Minkowski's inequality \cite{1988-Hardy-Inequalities}.
\section{Stability: Proof of Lemma 4.1}
\label{appendix:stability}
The MDPD $\PD(\MDRG_{\f})$ is computed by a restricted cartesian product of the points in $\PD(\RG_{f_1})$ and $\PD(\RG_{\widetilde{f_2^p}})$, where  $p$ varies in $\RG_{f_1}$. Let $\eta: \PD(\MDRG_{\f}) \rightarrow \PD(\MDRG_{\g})$ be the optimal bijection as required in the definition of $d_{W,q}(\MDRG_{\f},\MDRG_{\g})$. Let $\x = (a,b;c^p,d^p) \in \PD(\MDRG_{\f})$ and $\y = (a',b';c^{p'},d^{p'})\in \PD(\MDRG_{\g})$ be the matching pair based on $\eta$, i.e.  $\y=\eta(\x)$. Without loss of generality, let us consider $a \leq b, c^p \leq d^p, a' \leq b'$, and $c^{p'} \leq d^{p'}$ (i.e. we consider the points of the ordinary persistence diagrams). Then we have the following claim.

\begin{lemma}
\begin{footnotesize}
\begin{equation}
\label{eqn:claim}
\|\x - \eta(\x)\|_{\infty} \leq \max\{b-a, d^p-c^p, b'-a', d^{p'}-c^{p'}\}.
\end{equation}
\label{lemma:bound-l-infinity}
\end{footnotesize}
\end{lemma}
\begin{proof}
Assume that the  inequality (\eqnref{eqn:claim}) is not satisfied. Then we can match both $\x$ and $\y$ to their respective nearest diagonal points $Diag(\x) = (\frac{a+b}{2},\frac{a+b}{2};\frac{c^p+d^p}{2},\frac{c^p+d^p}{2})$ and $Diag(\y) = (\frac{a'+b'}{2},\frac{a'+b'}{2};\frac{c^{p'}+d^{p'}}{2},\frac{c^{p'}+d^{p'}}{2})$ to obtain the following bound:
\begin{footnotesize}
\begin{align*}
&\|\x - Diag(\x)\|_{\infty}^q + \|Diag(\y) - \y\|_{\infty}^q\\
&= \frac{1}{2^q} max\{(b-a)^q, (d^p-c^p)^q\} + \frac{1}{2^q}\max\{(b'-a')^q, (d^{p'}-c^{p'})^q\}\\
&\leq \frac{2}{2^q}\left(\max\{b-a, d^p-c^p, b'-a', d^{p'}-c^{p'}\}\right)^q\\
&=\frac{1}{2^{q-1}}\left(\max\{b-a, d^p-c^p, b'-a', d^{p'}-c^{p'}\}\right)^q\\
&\leq \left(\max\{b-a, d^p-c^p, b'-a', d^{p'}-c^{p'}\}\right)^q\\
& < \|\x - \y\|_{\infty}^q.
\end{align*}
\end{footnotesize}
This is a contradiction to the assumption that $\eta$ is the optimal bijection, since  $\x$ and $\y$ can be matched to their respective nearest diagonal points to obtain a lower value of $d_{W,q}$. 
\end{proof}

Let $\eta'$ be the restriction of $\eta$ on the non-diagonal points of $\PD(\MDRG_{\f})$. Then, for each point $\x= (a,b;c^p,d^p) \in \PD^{p}(\MDRG_{\f})$ (defined in \eqnref{eqn:PD^p}), we have $\eta'(\x) = (a',b'; c^{p'},d^{p'})$.
Thus we have
\begin{footnotesize}
\begin{align}
&\nonumber \sum_{\x \in \PD^{p}(\MDRG_{\f})}\|\x - \eta'(\x)\|_{\infty}^q \\
&\nonumber \leq \sum_{(a,b;c^p,d^p)\in \PD^{p}(\MDRG_{\f})} \left(\max\{b-a, d^p-c^p, b'-a', d^{p'}-c^{p'}\}\right)^q\\
& \nonumber \leq \sum_{(a,b;c^p,d^p)\in \PD^{p}(\MDRG_{\f})} \Bigl(\max\{Amp(f_1), Amp(\widetilde{f_2^p}),\\
&\hspace{4.5cm} Amp(g_1), Amp(\widetilde{g_2^{p'}})\}\Bigr)^q  \label{eqn:bound-PD_p}.
\end{align}
\end{footnotesize}
Further, from the construction of $\PD^{p}(\MDRG_{\f})$, it can be seen that the number of points in  $\PD^{p}(\MDRG_{\f})$ is at most the product of the number of points in $\PD(\RG_{f_1})$ and $\PD(\RG_{\widetilde{f_2^p}})$, i.e  $|\PD^{p}(\MDRG_{\f})| \leq |\PD(\RG_{f_1})| \cdot |\PD(\RG_{\widetilde{f_2^p}})|$. The \eqnref{eqn:bound-PD_p} can now be written as follows:
\begin{footnotesize}
\begin{equation*}
\begin{aligned}
  &\sum_{\x \in \PD^{p}(\MDRG_{\f})}\|\x - \eta'(\x)\|_{\infty}^q\\
  &\leq |\PD(\RG_{f_1})| \cdot |\PD(\RG_{\widetilde{f_2^p}})| \bigl(\max\{Amp(f_1), Amp(\widetilde{f_2^p}),\\
  &\hspace{4.5cm} Amp(g_1), Amp(\widetilde{g_2^{p'}})\}\bigr)^q.\\
  &\leq |\PD(\RG_{f_1})| \cdot |\PD(\RG_{\widetilde{f_2^p}})| \bigl(\max\{Amp(f_1), Amp(f_2),\\
  &\hspace{4.5cm}Amp(g_1), Amp(g_2)\}\bigr)^q.
\end{aligned}
\end{equation*}
\end{footnotesize}
We now obtain a bound on the term $\sum_{\x \in \PD(\MDRG_{\f})}\|\x - \eta'(\x)\|_{\infty}^q$ as follows:
\begin{footnotesize}
\begin{align*}
&\sum_{\x \in \PD(\MDRG_{\f})}\|\x - \eta'(\x)\|_{\infty}^q\\
&= \sum_{p \in \RG_{f_1}}\sum_{\x \in \PD^{p}(\MDRG_{\f})}\|\x - \eta'(\x)\|_{\infty}^q \\
&\leq\sum_{p \in \RG_{f_1}}|\PD(\RG_{f_1})| \cdot |\PD(\RG_{\widetilde{f_2^p}})|\Bigl(\max\{Amp(f_1), Amp(f_2),\\
&\hspace{5.3cm} Amp(g_1), Amp(g_2)\}\Bigr)^q\\
&\leq N_{\RG_{f_1}} |\PD(\RG_{f_1})| \max_{p \in \RG_{f_1}} |\PD(\RG_{\widetilde{f_2^p}})|\Bigl(\max\{Amp(f_1), Amp(f_2),\\
&\hspace{6cm}Amp(g_1), Amp(g_2)\}\Bigr)^q.
\end{align*}
\end{footnotesize}

\noindent Similar to the map $\eta'$, we define $\eta''$ as the restriction of $\eta$ on the non-diagonal points of $\PD(\MDRG_{\g})$. The following bound is obtained on  $\sum_{\x \in \PD(\MDRG_{\g})}\|{\eta''}^{-1}(\x) - \x\|_{\infty}^q$.

\begin{footnotesize}
\begin{equation*}
\begin{aligned}
&\sum_{\x \in \PD(\MDRG_{\g})}\|{\eta''}^{-1}(\x) - \x\|_{\infty}^q\\
&\leq N_{\RG_{g_1}} |\PD(\RG_{g_1})| \max_{p \in \RG_{g_1}} |\PD(\RG_{\widetilde{g_2^p}})|\Bigl( \max\{Amp(f_1), Amp(f_2),\\
&\hspace{6.1cm}Amp(g_1), Amp(g_2)\}\Bigr)^q.
\end{aligned}   
\end{equation*}
\end{footnotesize}

\noindent Thus we obtain a bound on the Wasserstein distance between $\PD(\MDRG_{\f})$ and $\PD(\MDRG_{\g})$ as follows:
\begin{footnotesize}
\begin{equation}
  \begin{aligned}
& d_{W,q}(\MDRG_{\f},\MDRG_{\g})\\
&\leq \Bigl( N_{\RG_{f_1}} |\PD(\RG_{f_1})| \max_{p \in \RG_{f_1}} |\PD(\RG_{\widetilde{f_2^p}})|\Bigl(\max\{Amp(f_1), Amp(f_2),\\
&\hspace{5.1cm} Amp(g_1), Amp(g_2)\}\Bigr)^q\\
&\hspace{0.4cm}+ N_{\RG_{g_1}} |\PD(\RG_{g_1})| \max_{p \in \RG_{g_1}} |\PD(\RG_{\widetilde{g_2^p}})| \Bigl(\max\{Amp(f_1), Amp(f_2),\\
&\hspace{5.1cm} Amp(g_1), Amp(g_2)\}\Bigr)^q\Bigr)^{\frac{1}{q}}.
\end{aligned}
\label{eqn:stability-bound-quantization}
\end{equation}
\end{footnotesize}

\noindent Finally, we bound the number of points in $\PD(\RG_{f})$ and $\PD(\RG_{g})$ by the number of critical points of the corresponding functions. We note, each point in a persistence diagram  $\PD(\RG_{f})$ corresponds to a pair of birth-death events which occur at critical points of the function $\bar{f}$, defined on $\RG_{f}$. Therefore, the number of points in $\PD(\RG_{f})$ is at most the number of critical points of $\bar{f}$, which is upper bounded by the number of critical points of $f$. The \eqnref{eqn:stability-bound-quantization} can now be written as follows.
\begin{footnotesize}
   \begin{align*}
  &d_{W,q}(\MDRG_{\f}, \MDRG_{\g})\\
&\leq\biggl( N_{\RG_{f_1}} C_{f_1} \max_{p \in \RG_{f_1}} C_{\widetilde{f_2^p}} \Bigl(\max\{Amp(f_1), Amp(f_2),\\
&\hspace{4.5cm}Amp(g_1),Amp(g_2)\}\Bigr)^q\\
&\hspace{0.2cm}+N_{\RG_{g_1}} C_{g_1} \max_{p \in \RG_{g_1}} C_{\widetilde{g_2^p}} \Bigl(\max\{Amp(f_1),Amp(f_2),\\
&\hspace{4.5cm}Amp(g_1),Amp(g_2)\}\Bigr)^q\biggr)^{\frac{1}{q}}
   \end{align*} 
\end{footnotesize}

\noindent Thus we obtain a bound on $d_{W,q}$ based on the number of nodes in the quantized Reeb graphs $\RG_{f_1}, \RG_{g_1}$, the number of critical points of $f_1, g_1, \widetilde{f_2^p}, \widetilde{g_2^p}$, and the amplitudes of $f_1, f_2, g_1$ and $g_2$.
\section{Generalization for More Than Two Fields}
\label{appendix:generalization-MDPD-n-fields}
In this paper, we deal with MDPDs of bivariate fields. However, the definition of MDPD can be generalized for more than $2$ fields. Let $\f = (f_1, f_2, \ldots, f_n) : \cM \rightarrow \R^n$ be a multi-field defined on a compact $m$-manifold $\cM$ with $m \geq n \geq 2$. We then define the MDPD corresponding to $\MDRG_{\f}$ by generalizing \eqnref{eqn:MDPD-2-fields-full-version}, as follows.
\begin{footnotesize}
\begin{eqnarray}
&\PD(\MDRG_\f) :=\displaystyle  \bigcup_{\ux_1\in\PD(\RG_{f_1})} \bigcup_{\left\{p_1\in \RG_{f_1}: \bar{f}_1(p_1)\in \pI(\ux_1)\right\}} \nonumber\\
& \displaystyle  \bigcup_{\ux_2^{p_1} \in \PD(\RG_{\widetilde{f_2^{p_1}}})} \bigcup_{\left\{p_2 \in \RG_{\widetilde{f_2^{p_1}}}\bigg\vert \overline{\widetilde{f_2^{p_1}}}(p_2) \in \pI\left(\ux_2^{p_1}\right)\right\}} \nonumber   \\
& \displaystyle \bigcup_{\ux_3^{p_2} \in \PD(\RG_{\widetilde{f_3^{p_2}}})} \bigcup_{\left\{p_3 \in \RG_{\widetilde{f_3^{p_2}}}\bigg\vert \overline{\widetilde{f_3^{p_2}}}(p_3) \in \pI\left(\ux_3^{p_2}\right)\right\}}\nonumber \\      
&\vdots \nonumber\\
& \displaystyle \bigcup_{\ux_{n-1}^{p_{n-2}} \in \PD(\RG_{\widetilde{f_{n-1}^{p_{n-2}}}})} \bigcup_{\left\{p_{n-1} \in \RG_{\widetilde{f_{n-1}^{p_{n-2}}}} \bigg\vert \overline{\widetilde{f_{n-1}^{p_{n-2}}}}(p_{n-1}) \in \pI\left(\ux_{n-1}^{p_{n-2}}\right)\right\}}\nonumber\\
& \displaystyle \bigcup_{\ux_n^{p_{n-1}} \in \PD(\RG_{\widetilde{f_n^{p_{n-1}}}})} (\ux_1; \ux_2^{p_1};\ux_3^{p_2}; \ldots ; \ux_n^{p_{n-1}})
\label{eqn:MDPD-n-fields}
\end{eqnarray}
\end{footnotesize}
which is a multiset of points in $\R^2 \times \R^2 \times \ldots \times \R^2$ ($n$-times) or $\R^{2n}$. We note, a point $(\ux_1; \ux_2^{p_1};\ux_3^{p_2}; \ldots ; \ux_n^{p_{n-1}}) \in\PD(\MDRG_{\f})$ represents the persistence of $n$ homology classes $(\gamma_1, \gamma_2^{p_1},\gamma_3^{p_2}, \ldots , \gamma_n^{p_{n-1}})$ corresponding to the filtrations of $\RG_{f_1}, \RG_{\tilde{f_2^{p_1}}}, \RG_{\tilde{f_3^{p_2}}}, \ldots, \RG_{\tilde{f_n^{p_{n - 1}}}}$, respectively. The persistence measure of $(\gamma_1, \gamma_2^{p_1},\gamma_3^{p_2}, \ldots, \gamma_n^{p_{n-1}})$ is defined as the product of the lengths of the persistence intervals $\pI(\ux_1), \pI(\ux_2^{p_1}), \pI(\ux_3^{p_2}), \ldots \pI(\ux_n^{p_{n - 1}})$. Geometrically, it represents the volume of the orthotope given by the cartesian product $\pI(\ux_1) \times \pI(\ux_2^{p_1}) \times \pI(\ux_3^{p_2}) \times \ldots \pI(\ux_n^{p_{n - 1}})$.

\subsection{Distance between MDPDs}
Let $\f = (f_1, f_2, \ldots, f_n)$ and $\g = (g_1,g_2, \ldots, g_n)$ be two multi-fields defined on a compact $m$-manifold such that $m \geq n \geq 2$. Let $\JCN_{\f}$ and $\JCN_{\g}$ be the corresponding JCNs having identical quantization levels, and $\MDRG_{\f}, \MDRG_{\g}$ be the corresponding MDRGs. Then the definition of $d_{W,q}(\MDRG_{\f}, \MDRG_{\g})$ follows from \eqnref{eqn:Wasserstein-Distance-MDPDs}, with the maps $\eta : \PD(\MDRG_{\f}) \rightarrow \PD(\MDRG_{\g})$ satisfying the following criteria.

\begin{enumerate}[label=(C{{\arabic*}})]
    \item \textbf{Matching points from the same levels for each of the component fields:}  Let $\pointa = (a_1,b_1;a_2^{p_1},b_2^{p_1};\ldots;a_n^{p_{n-1}},b_n^{p_{n-1}}) \in \PD(\MDRG_{\f})$. Then $\eta$ maps $\pointa$ either to a point $\pointb=(c_1,d_1;c_2^{p_1'},d_2^{p_1'};\ldots;c_n^{p_{n-1}'}, d_n^{p_{n-1}'}) \in \PD(\MDRG_{\g})$ such that $\bar{f_1}(p_1) = \bar{g_1}(p_1'), \overline{\widetilde{f_2^{p_1}}}(p_2) = \overline{\widetilde{g_2^{p_1'}}}(p_2'),$ \ldots $, \overline{\widetilde{f_{n-1}^{p_{n-2}}}}(p_{n-1}) = \overline{\widetilde{g_{n-1}^{p_{n-2}'}}}(p_{n-1}'),$ or $\eta(\pointa)$ is a diagonal point of $\PD(\MDRG_{\g})$. This condition is imposed to ensure that points in the MDPDs are matched by $\eta$ only when they correspond to the same levels for each of the first $n - 1$ component functions of $\f$ and $\g$.
     
    \item \textbf{Topological consistency:} If  $\pointa = (a_1,b_1;a_2^{p_1},b_2^{p_1};\ldots;a_n^{p_{n-1}},b_n^{p_{n-1}})  \in \PD(\MDRG_{\f})$ is matched with  $\pointb = (c_1,d_1;c_2^{p_1'},d_2^{p_1'};\ldots;c_n^{p_{n-1}'}, d_n^{p_{n-1}'}) \in \PD(\MDRG_{\g})$ by $\eta$, then for all other points of the form $\x=(a_{i1}, b_{i1}; a_{i2}^{p_1}, b_{i2}^{p_1}; \ldots; a_{in}^{p_{n-1}}, b_{in}^{p_{n-1}})\in \PD(\MDRG_{\f})$, $\eta(\x)$ will be of the form $(c_{i1}, d_{i1}; c_{i2}^{p_1'}, d_{i2}^{p_1'}; \ldots; c_{in}^{p_{n-1}'}, d_{in}^{p_{n-1}'})\in \PD(\MDRG_{\g})$, or a diagonal point of $\PD(\MDRG_{\g})$. That is, all the points in $\PD(\MDRG_{\f})$ obtained from a persistence diagram $\PD(\RG_{\widetilde{f_i^{p_{i-1}}}})$ are mapped to points in $\PD(\MDRG_{\g})$ obtained from the persistence diagram $\PD(\RG_{\widetilde{g_i^{p_{i-1}'}}})$. In other words, no two points in $\PD(\MDRG_{\f})$ coming from $\PD(\RG_{\widetilde{f_i^{p_{i-1}}}})$ are mapped to points in $\PD(\MDRG_{\g})$ coming from two different persistence diagrams $\PD(\RG_{\widetilde{g_i^{p_{i-1}'}}})$ and $\PD(\RG_{\widetilde{g_i^{p_{i-1}''}}})$ with $p_{i-1}' \neq p_{i-1}''$. This condition checks for topological consistency by ensuring that for $1 \leq i \leq n-1$, all points in $\PD(\MDRG_{\f})$ corresponding to a contour of $f_i$ are mapped to points in $\PD(\MDRG_{\g})$ corresponding to a contour of $g_i$.

\item \textbf{Dimension consistency:} $\pointa=(a_1,b_1;a_2^{p_1},b_2^{p_1};\ldots;a_n^{p_{n-1}},b_n^{p_{n-1}}) \in \PD(\MDRG_{\f})$ is matched with $\pointb=(c_1,d_1;c_2^{p_1'},d_2^{p_1'};\ldots;c_n^{p_{n-1}'}, d_n^{p_{n-1}'}) \in \PD(\MDRG_{\g})$ by $\eta$ if (i) the dimensions of the persistent homology classes corresponding to $(a_1,b_1)$ and $(c_1,d_1)$ are the same, and (ii) the dimensions of the persistent homology classes corresponding to $(a_i^{p_{i-1}}, b_i^{p_{i-1}})$ and $(c_i^{p_{i-1}'}, d_i^{p_{i-1}'})$ are the same, for $2 \leq i \leq n$.
\end{enumerate}

\subsection{Properties}
\textbf{Pseudo-metric:} Similar to the case of bivariate fields, $d_{W,q}(\MDRG_{\f}, \MDRG_{\g})$ satisfies the properties of a pseudo-metric. The proof of this is similar to that of \thmref{theorem:pseudo-metric}.

\noindent \textbf{Stability:} From the construction of $\PD(\MDRG_{\f})$, we obtain the following bound on the number of points it contains.

\begin{footnotesize}
\begin{align}
&|\PD(\MDRG_{\f})| \nonumber\\
&\leq \Bigg(|\PD(\RG_{f_1})|N_{\RG_{f_1}} \max_{p_1 \in \RG_{f_1}} |\PD(\RG_{\widetilde{f_2^{p_1}}})|\\
&\hspace{0.8cm} N_{\RG_{\widetilde{f_2^{p_1}}}} \max_{p_2 \in \RG_{\widetilde{f_2^{p_1}}}} |\PD(\RG_{\widetilde{f_3^{p_2}}})|\ldots\nonumber\\
&\hspace{0.6cm}\ldots N_{\RG_{\widetilde{f_{n-1}^{p_{n-2}}}}} \max_{p_{n-1} \in \RG_{\widetilde{f_{n-1}^{p_{n-2}}}}} |\PD(\RG_{\widetilde{f_n^{p_{n-1}}}})|  \Bigg)\nonumber\\
&\leq  \Bigg(C_{f_1} N_{\RG_{f_1}} \max_{p_1 \in \RG_{f_1}} C_{\widetilde{f_2^{p_1}}} N_{\RG_{\widetilde{f_2^{p_1}}}} \max_{p_2 \in \RG_{\widetilde{f_2^{p_1}}}} C_{\widetilde{f_3^{p_2}}}  \ldots \nonumber\\
&\hspace{0.6cm}\ldots N_{\RG_{\widetilde{f_{n-1}^{p_{n-2}}}}}\max_{p_{n-1} \in \RG_{\widetilde{f_{n-1}^{p_{n-2}}}}} C_{\widetilde{f_n^{p_{n-1}}}}  \Bigg). \label{eqn:stability-n-fields-bound1}
\end{align}
\end{footnotesize}

\noindent Similarly, the number of points in $\PD(\MDRG_{\g})$ is bounded above as follows.
\begin{footnotesize}
\begin{align}
&|\PD(\MDRG_{\g})|\nonumber\\
&\leq \Bigg(C_{g_1} N_{\RG_{g_1}} \max_{p_1 \in \RG_{g_1}} C_{\widetilde{g_2^{p_1}}} N_{\RG_{\widetilde{g_2^{p_1}}}} \max_{p_2 \in \RG_{\widetilde{g_2^{p_1}}}} C_{\widetilde{g_3^{p_2}}}  \ldots  \nonumber\\
&\hspace{0.6cm}\ldots N_{\RG_{\widetilde{g_{n-1}^{p_{n-2}}}}}\max_{p_{n-1} \in \RG_{\widetilde{g_{n-1}^{p_{n-2}}}}} C_{\widetilde{g_n^{p_{n-1}}}}  \Bigg). \label{eqn:stability-n-fields-bound2}
\end{align}
\end{footnotesize}
\noindent Let $\x = (a_1,b_1;a_2^{p_1},b_2^{p_1};a_3^{p_2},b_3^{p_2}, \ldots, a_n^{p_{n-1}}, b_n^{p_{n-1}}) \in \PD(\MDRG_{\f})$ and $\eta(\x) = (c_1,d_1;c_2^{p_1'},d_2^{p_1'};c_3^{p_2'},d_3^{p_2'}, \ldots, c_n^{p_{n-1}'}, d_n^{p_{n-1}'})$. Then the following bound is obtained on $\|\x - \eta(\x)\|_{\infty}$ (similar to  Lemma \ref{lemma:bound-l-infinity}).

\begin{footnotesize}
\begin{align}
\|\x - \eta(\x)\|_{\infty} &\leq  \max\{b_1 - a_1, d_1 - c_1, b_2^{p_1} - a_2^{p_1}, d_2^{p_1'} - c_2^{p_1'}, \ldots \nonumber\\
&\hspace{1.2cm} \ldots, b_n^{p_{n-1}} - a_n^{p_{n-1}}, d_n^{p_{n-1}'} - c_n^{p_{n-1}'}\} \nonumber\\
&\hspace{0.4cm}\leq \max \left\{\max_{1 \leq i \leq n} Amp(f_i), \max_{1\leq i \leq n} Amp(g_i)\right\}. \label{eqn:stability-n-fields-bound3}
\end{align}
\end{footnotesize}

\noindent Let $T_1$ and $T_2$ denote the RHS of equations (\ref{eqn:stability-n-fields-bound1}) and (\ref{eqn:stability-n-fields-bound2}), respectively. Then we obtain the following bound on $d_{W,q}(\MDRG_{\f},\MDRG_{\g})$, based on equations (\ref{eqn:stability-n-fields-bound1}), (\ref{eqn:stability-n-fields-bound2}) and (\ref{eqn:stability-n-fields-bound3}).

\begin{theorem}
\begin{footnotesize}
\begin{equation*}
\begin{aligned}
&d_{W,q}(\MDRG_{\f}, \MDRG_{\g})\\
&\leq\left(\left( T_1 + T_2\right) \left(\max \left\{\max_{1 \leq i \leq n} Amp(f_i), \max_{1\leq i \leq n} Amp(g_i)\right\}\right)^q\right)^{\frac{1}{q}}.
\end{aligned}
\end{equation*}
\label{thm:stability-n-fields}
\end{footnotesize}
\end{theorem}

\subsection{Complexity Analysis}
In this subsection, we analyze the complexities of constructing the MDPD from the MDRG of a multi-field and the distance between two MDRGs based on their MDPDs.

Let $\f = (f_1, f_2, \ldots, f_n): \cM \rightarrow \R^n$ be a multi-field defined on a compact $m$-manifold  $\cM$ with $m \geq n \geq 2$. The time complexity of computing $\MDRG_{\f}$ from $\JCN_{\f}$ is $\cO(n|V|(|V| + |E| \alpha(|V|) + |V| \log(|V|)))$, where $|V|$ and $|E|$ are number of vertices and edges, respectively, in $\JCN_{\f}$. The total number of simplices (vertices and edges) corresponding to all the Reeb graphs of any particular dimension of $\MDRG_{\f}$, is at most $\cO(|V| + |E|)$. The time complexity for constructing the persistence diagrams of these Reeb graphs is $\cO((|V| + |E|)^3)$. Thus, the total time for constructing the persistence diagrams of the Reeb graphs in all the dimensions of $\MDRG_{\f}$ is $\cO(n(|V| + |E|)^3)$, as discussed in \cite{book-herbert-computational-topology}. After the persistence diagrams of the component Reeb graphs in $\MDRG_{\f}$ are computed, it takes $(|V|^n)$ time to compute the MDPD (similar to the analysis in \secref{subsubsec:complexity-mdpd-construction}). The total time for the construction of $\PD(\MDRG_{\f})$ is $\cO(n|V|(|V| + |E| \alpha(|V|) + |V| \log(|V|)) + n(|V| + |E|)^3 + |V|^n) \sim \cO(n(|V| + |E|)^3 + |V|^n)$.

Next, we discuss the time complexity for computing the distance between two MDPDs. Let $\f = (f_1, f_2, \ldots, f_n)$ and $\g = (g_1, g_2, \ldots, g_n)$ be two multi-fields defined on a compact $m$-manifold  $\cM$ with $m \geq n \geq 2$. Let $\PD(\MDRG_{\f})$ and $\PD(\MDRG_{\g})$ be the corresponding MDPDs. We note, $\PD(\MDRG_{\f})$ is constructed by taking a restricted cartesian product of the points in $\PD(\RG_{f_1}), \PD(\RG_{\widetilde{f_2^{p_1}}}), \PD(\RG_{\widetilde{f_3^{p_2}}}), \ldots, \PD(\RG_{\widetilde{f_n^{p_{n - 1}}}})$, where $p_1 \in \RG_{f_1}$ and $p_i \in \RG_{\widetilde{f_i^{p_{i-1}}}}$ for $2 \leq i \leq n - 1$. Further, the number of points in the persistence diagrams of the quantized Reeb graph $\RG_{f}$ is at most the number of critical points of $\bar{f}$, which is bounded above by the number of simplices in $\RG_{f}$. Further, the total number of vertices in the Reeb graphs belonging to any particular dimension of $\MDRG_{\f}$ is bounded above by the number of JCN vertices, i.e. $|V_1|$. Therefore, the number of points in $\PD(\MDRG_{\f})$ is at most $|V_1|^{n}$, where $|V_1|$ is the number of vertices in $\JCN_{\f}$. Similarly, the number of points in $\PD(\MDRG_{\g})$ is bounded above by $|V_2|^n$, where $|V_2|$ is the number of vertices in $\JCN_{\g}$. To compute $d_{W,q}(\MDRG_{\f}, \MDRG_{\g})$, we consider bijections between $\PD(\MDRG_{\f})$ and $\PD(\MDRG_{\g})$ satisfying the criteria (C1) - (C3). However, in the worst case, when we consider all possible bijections between the MDPDs, the time complexity of computing the Wasserstein distance between the MDPDs using the Hungarian algorithm \cite{1955-Kuhn-Hungarian-Algorithm} is $\cO((|V_1|^n  + |V_2|^n)^3)$.

\end{document}